\documentclass[12pt,draftcls,journal,twoside,onecolumn,romanappendices]{IEEEtran}
\usepackage[tbtags]{amsmath}
\usepackage{amsmath}

\usepackage{algorithm}
\usepackage{algpseudocode}

\usepackage{amssymb}
\usepackage{cite}
\usepackage{stfloats}

\usepackage{amsthm}
\usepackage{float}
\usepackage{amsthm}
\usepackage{amssymb}

\usepackage{graphicx}%
\usepackage{subfigure}
\usepackage{acronym}
\usepackage[keeplastbox]{flushend}
\usepackage{stfloats}
\usepackage{color}

\newcommand{\mbf}[1]{\mathbf{#1}}
\newcommand{\ub}[1]{\underbrace{#1}}

\newtheoremstyle{myremark}% 
  {}%                   Space above, empty = `usual value'
  {}%                   Space below
  {}%                   Body font
  {\parindent}%         Indent amount (empty = no indent, \parindent = para indent)
  {\itshape}%          Thm head font
  {:}%                   Punctuation after thm head
  {5pt plus 1pt minus 1pt}%                Space after thm head
  {\thmname{#1}\thmnumber{~#2}\thmnote{~(#3)}}%                   Thm head spec

\theoremstyle{myremark}
\newtheorem{theorem}{Theorem}
\newtheorem{corollary}{Corollary}
\newtheorem{lemma}{Lemma}
\theoremstyle{myremark}
\newtheorem*{remark}{Remark}

\acrodef{CRN}{cognitive radio network}
\acrodef{SU}{secondary user}
\acrodef{PU}{primary user}
\acrodef{ZF}{zero forcing}
\acrodef{FD}{full-duplex}
\acrodef{BS}{base station}
\acrodef{i.i.d.}{independent and identically distributed}
\acrodef{DL}{downlink}
\acrodef{UL}{uplink}
\acrodef{SINR}{signal-to-interference-plus-noise ratio}
\acrodef{SNR}{signal noise ratio}
\acrodef{AWGN}{additive white Gaussian noise}
\acrodef{MMSE}{minimum mean square error}
\acrodef{SIC}{successive interference cancellation}
\acrodef{SI}{self-interference}
\acrodef{CCI}{co-channel interference}
\acrodef{MUI}{multiuser interference}
\acrodef{NOMA}{non-orthogonal multiple access}
\acrodef{OMA}{orthogonal multiple access}
\acrodef{QoS}{quality of service}
\acrodef{SIC}{successive interference cancellation}
\acrodef{SVD}{singular value decomposition}
\acrodef{MIMO}{multiple-input multiple-output}
\acrodef{SISO}{single-input single-output}
\acrodef{MIMO-NOMA}{multiple-input multiple-output non-orthogonal multiple access}
\acrodef{MIMO-OMA}{multiple-input multiple-output orthogonal multiple access}

\begin{document}
%
% paper title
% Titles are generally capitalized except for words such as a, an, and, as,
% at, but, by, for,  in, nor, of, on, or, the, to and up, which are usually
% not capitalized unless they are the first or last word of the title.
% Linebreaks \\ can be used within to get better formatting as desired.
% Do not put math or special symbols in the title.
\title{Capacity Comparison between MIMO-NOMA and MIMO-OMA with Multiple Users in a Cluster}
%\\ for IEEE Conferences}

%\author{\IEEEauthorblockN{Ming Zeng, \emph{Student Member, IEEE},
%Animesh Yadav, \emph{Member, IEEE}, Octavia A. Dobre, \emph{Senior Member}, \emph{IEEE}, Georgios I. Tsiropoulos, \emph{Member, IEEE}, and \\H. Vincent Poor, \emph{Fellow, IEEE} 
\author{\IEEEauthorblockN{Ming Zeng,
Animesh Yadav,  Octavia A. Dobre,  Georgios I. Tsiropoulos \\and H. Vincent Poor
%\IEEEauthorblockA{\IEEEauthorrefmark{1}Faculty of Engineering and Applied Science, Memorial University, St. John, Canada}
%\IEEEauthorblockA{\IEEEauthorrefmark{2}School of Electrical and Computer Engineering, National Technical University of Athens, Greece}
}

%\IEEEauthorblockA {Email: mzeng@mun.ca}}
%\IEEEauthorblockA {Email: mzeng, animeshy, odobre@mun.ca, gitsirop@ntua.gr,  and poor@princeton.edu} 
}

\maketitle

% As a general rule, do not put math, special symbols or citations
\vspace*{-2.5\baselineskip}

\begin{abstract}
In this paper, the performance of \ac{MIMO-NOMA} is investigated when multiple users are grouped into a cluster. The superiority of \ac{MIMO-NOMA} over MIMO orthogonal multiple access (\acs{MIMO-OMA}) in terms of both {\color{black}sum channel capacity} and {\color{black}ergodic sum capacity} is proved analytically. Furthermore, it is demonstrated that the more users are admitted to a cluster, the lower is the achieved {\color{black}sum rate}, which illustrates the tradeoff between the {\color{black}sum rate} and maximum number of admitted users. On this basis, a user admission scheme is proposed, which is optimal in terms of both {\color{black}sum rate} and number of admitted users when the \acl{SINR} thresholds of the users are equal. When these thresholds are different, the proposed scheme still achieves good performance in balancing both criteria. Moreover, under certain conditions, it maximizes the number of admitted users. In addition, the complexity of the proposed scheme is linear to the number of users per cluster. Simulation results verify the superiority of \ac{MIMO-NOMA} over \acs{MIMO-OMA} {\color{black}in terms of both sum rate and user fairness}, as well as the effectiveness of the proposed user admission scheme. 
\end{abstract}

%\vspace*{-1.5\baselineskip}
%\begin{IEEEkeywords}
%Non-orthogonal multiple access (NOMA), multiple-input multiple-output (MIMO), channel capacity, sum rate, fairness, user admission, power allocation.
%\end{IEEEkeywords}
{\let\thefootnote\relax\footnote{This research was supported in part by the Natural Sciences and Engineering Research Council of Canada (NSERC) through its Discovery program and the U.S. National Science Foundation under Grants CCF-1420575 and CNS-1456793.

O. A. Dobre, A. Yadav, and M. Zeng are with Memorial University, St. John's, NL A1B 3X9, Canada (e-mail: odobre, animeshy, mzeng@mun.ca).

G. I. Tsiropoulos is with National Technical University of Athens, 10682, Athens, Greece (e-mail: gitsirop@mail.ntua.gr).

H. V. Poor is with Princeton University, Princeton, NJ 08544 USA (e-mail: poor@princeton.edu).}  }
\IEEEpeerreviewmaketitle

\vspace*{-1\baselineskip}

\section{Introduction}

\acused{NOMA} Non-orthogonal multiple access (NOMA) has attracted considerable attention recently due to its superior spectral efficiency \cite{1, 2, 3, 12, 15, 16, 20}. Specifically, \ac{NOMA} adopts superposition coding (SC) at the transmitter and \ac{SIC} at the receiver. Moreover, the transmitted power allocated to the users is inversely proportional to their channel gains. This way, the user with better channel gain can handle the interference from its counterpart, while its interference to the counterpart remains comparatively small. Thus, \ac{NOMA} achieves a better balance between {\color{black}sum rate} and fairness when compared with conventional \ac{OMA} scheme, in which more power is assigned to the users with better channel conditions to increase the {\color{black}sum rate} \cite{13}.

It is of great interest to conduct comparisons between \ac{NOMA} and \ac{OMA}. Early works mainly focus on \ac{SISO} systems. For instance, simulation results in \cite{1} show that a larger {\color{black}sum rate} is achieved by \ac{NOMA}, whereas in \cite{4}, it is proved that \ac{NOMA} strictly dominates \ac{OMA} via the achievable rate region. However, no analytical proof is provided in \cite{1} and \cite{4}. In \cite{9}, the performance of NOMA is investigated in a cellular downlink scenario with randomly deployed users, and the developed analytical results show that NOMA can achieve superior performance in terms of ergodic {\color{black}sum rate}. In \cite{13}, the problem of maximizing the fairness among users of a NOMA downlink system is studied in terms of data rate under full channel state information (CSI) and outage probability under average CSI. Simulation results verify the efficiency of NOMA, which also achieves improved fairness when compared to time division multiple access.

%Recently, more efforts are made to 
Emerging research activities in future mobile wireless networks study the performance of \ac{NOMA} under \ac{MIMO} channels. In \cite{5}, the authors explore the two user power allocation problem of a NOMA scheme by maximizing the ergodic {\color{black}sum} capacity of MIMO channel under the total transmit power, minimum rate requirement and partial CSI availability constraints. Optimal and lower complexity power allocation schemes are proposed, and numerical results show that \ac{MIMO}-\ac{NOMA} obtains a larger ergodic sum {\color{black}capacity} when compared to \ac{MIMO}-\ac{OMA}. In \cite{6,8}, Ding \text{et al.} investigate the performance of \ac{MIMO}-\ac{NOMA} when there are multiple clusters in the system and, through simulations, validate the superiority of \ac{MIMO}-\ac{NOMA} over \ac{MIMO}-\ac{OMA}. Specifically, \cite{6} studies the \ac{DL} with limited feedback at the \ac{BS}, while \cite{8} considers both \ac{DL} and uplink with full CSI at the user side and \ac{BS}. Additionally, for each cluster, multiple users can be admitted into \cite{6}, whereas \cite{8} can only support two users performing signal alignment. However, neither \cite{6} nor \cite{8} provides an analytical comparison between \ac{MIMO}-\ac{NOMA} and \ac{MIMO}-\ac{OMA} in terms of {\color{black}sum rate}. Based on the system model proposed in \cite{6}, \cite{7} conducts the {\color{black}sum rate} comparison between them when there are only two users in each cluster. It is shown analytically that for any rate pair achieved by \ac{MIMO}-\ac{OMA}, there is a power split for \ac{MIMO}-\ac{NOMA} whose rate pair is larger. Despite the attractiveness of the result, its main issue is that the authors use the Jensen's inequality and concavity of $\log(\cdot)$ inappropriately to obtain the upper bound {\color{black}sum rate} for \ac{MIMO}-\ac{OMA}. {\color{black}In \cite{18}, it is shown that for a simple scenario of two users, MIMO-NOMA dominates MIMO-OMA in terms of {\color{black}sum rate}. Furthermore, for a more practical scenario of multiple users, with two users paired into a cluster and sharing a common transmit beamforming vector, the conclusion still holds.}

Most of the existing works in MIMO-NOMA focus on the case of two users in each cluster \cite{3,5, 6,7,8, 17, 18}, which leads to a less-studied alternative in the case of multiple users \cite{6, 14}. In order to serve more users simultaneously, it is of great significance to investigate the performance of \ac{MIMO}-\ac{NOMA} with multiple users per cluster. Although \cite{6} can support multiple users per cluster, the authors focus on user pairing and power allocation for the two user case. In \cite{14}, the proposed \ac{MIMO}-\ac{NOMA} scheme requires only one-bit feedback, but power allocation is not addressed, and there is no theoretical comparison of the performance of \ac{MIMO}-\ac{NOMA} and \ac{MIMO}-\ac{OMA}. In this paper, we aim to analytically compare the performance of \ac{MIMO-NOMA} with \acs{MIMO-OMA} in terms of the {\color{black}sum channel capacity} and {\color{black}ergodic sum capacity} rather than merely providing simulation results, when there are multiple users in a cluster. Furthermore, the study of the way the {\color{black}sum rate} varies as the number of admitted users increases in each cluster is conducted. To the best of our knowledge, this paper is the first to address this issue under \ac{MIMO-NOMA} systems. Following this, optimal user admission is investigated in terms of the number of admitted users and {\color{black}sum rate}, when the target \acf{SINR} of each user is given. Compared with the existing works, the main contribution of this paper lies in:

\begin{itemize}
  \item We prove analytically that \ac{MIMO}-\ac{NOMA} outperforms \ac{MIMO}-\ac{OMA} in terms of both {\color{black}sum channel capacity} and {\color{black}ergodic sum capacity} when there are multiple users in a cluster. We show that for any power split in \ac{MIMO}-\ac{OMA}, a larger {\color{black}sum rate} can be achieved by \ac{MIMO}-\ac{NOMA} via simply assigning the same power coefficient to the latter. In addition, for the case of two users per cluster, we derive the power split that maximizes the {\color{black}sum rate} gap between \ac{MIMO-NOMA} and \acs{MIMO-OMA}. {\color{black}Meanwhile, numerical results validate that \ac{MIMO}-\ac{NOMA} also achieves higher user fairness than \ac{MIMO}-\ac{OMA} when there are two or three users in a cluster.}
%  \item It is validated that \ac{SIC} at the receiver side is guaranteed to be successful due to the channel ordering in \ac{MIMO-NOMA}, i.e., user with better channel condition can remove the interference from those with worse channel conditions. 
  \item We demonstrate that as more users are admitted to a cluster, the {\color{black}sum rate} decreases. This illustrates that a tradeoff has to be considered between the {\color{black}sum rate} and number of admitted users. On this basis, we propose a user admission scheme, which aims to maximize the number of admitted users under given \ac{SINR} thresholds. The proposed scheme is shown to be optimal when the \ac{SINR} thresholds for users in the same cluster are equal. Otherwise, it achieves a good balance between the {\color{black}sum rate} and number of admitted users. Furthermore, under certain conditions, the proposed scheme maximizes the number of admitted users.  Additionally, its complexity is linear. 
\end{itemize}

The rest of the paper is organized as follows. The system model is introduced in Section II. In Section III, the capacity comparison between \ac{MIMO}-\ac{NOMA} and \ac{MIMO}-\ac{OMA} is conducted. The proposed user admission scheme is introduced in Section IV, while simulation results are shown in \text{Section V}. In Section VI, conclusions are drawn.

%Similar to the existing work in \cite{6,7}, we assume that $N \geq M$, in order to eliminate the interference amongst clusters. In addition, Following the design of precoding matrices and detection vectors in \cite{6

\section{System Model}
A downlink multiuser MIMO system is considered in this paper, where the \ac{BS} with $M$ antennas transmits data to multiple receivers, each with $N$ antennas. There are a total of $ML$ users in the system, which are randomly grouped into $M$ clusters with $L\: (L \geq 2)$ users per cluster. The links between the BS and users are assumed to be quasi-static independent and identically distributed (i.i.d.) fading channels. Specifically, $\mbf{H}_{m,l}\in \mathbb{C}^{N\times M}$ and $\mbf{n}_{m,l}\in \mathbb{C}^{N \times 1}$ respectively represent the channel matrix and the additive white Gaussian noise vector for the $l$th user in the $m$th cluster, i.e., user $(m,l)$ $(m\in\{1,\dots,M\}, l\in\{1,\dots,L\})$. Additionally, $\mbf{P}\in \mathbb{C}^{M\times M}$ denotes the precoding matrix used by the BS, while $\mbf{v}_{m,l}\in \mathbb{C}^{N\times 1}$ denotes the detection vector for user $(m,l)$. The precoding matrices and detection vectors are designed as follows \cite{6}: a) $\mbf{P}=\mbf{I}_M$, where $\mbf{I}_M$ denotes the $M \times M$ identity matrix; b) $|\mbf{v}_{m,l}|^2=1$ and ${\mbf{v}_{m,l}^H}\mbf{H}_{m,l}\mbf{p}_k=0$ for any $k\neq m$, where $\mbf{p}_k$ is the $k$th column of $\mbf{P}$.  \textcolor{black}{The number of antennas at the user is assumed to be equal or larger than that at the \ac{BS} to ensure the feasibility of $\mathbf{v}_{m,l}$.} On this basis, for user $(m,l)$, only a scalar value $|{\mbf{v}_{m,l}^H}\mbf{H}_{m,l}\mbf{p}_m|^2$ needs to be fed back to the \ac{BS}. Moreover, the interference from the users in all the other clusters can be removed even when there are multiple users in a cluster\cite{6}. 

%which follow any distribution, e.g., Rayleigh distribution

The performance of two multiple access schemes are compared, namely, \ac{MIMO}-\ac{NOMA} and \ac{MIMO}-\ac{OMA}.

\subsection{\ac{MIMO}-\ac{NOMA}}
For \ac{MIMO}-\ac{NOMA} scheme, SC is employed at the transmitter side, i.e., the transmitted signals share the same frequency and time resources but vary in power. Thus, the signals transmitted from the \ac{BS} are given by
\begin{equation}
\mbf{x}=\mbf{P}\mbf{s},
\end{equation}
where the information-bearing vector $\mbf{s} \in \mathbb{C}^{M\times 1}$ can be expressed as
\begin{equation}
\mbf{s}=
\begin{bmatrix}
\sqrt{\Omega_{1,1}}s_{1,1}+\dots + \sqrt{\Omega_{1,L}}s_{1,L} \\
\vdots \\
\sqrt{\Omega_{M,1}} s_{M,1}+\dots +\sqrt{\Omega_{M,L}}s_{M,L}
\end{bmatrix},
\end{equation}
where $s_{m,l}$ and $\Omega_{m,l}$ are the signal and the corresponding power allocation coefficient intended for user $(m,l)$, satisfying $\sum_{l=1}^{L} \Omega_{m,l}=1, \forall m\in\{1,\dots,M\}$. Without loss of generality, we set the total power to 1 for the convenience of analysis. 

Further, the received signal at user $(m,l)$ is given by
\begin{equation}
\mbf{y}_{m,l}=\mbf{H}_{m,l}\mbf{P}\mbf{s}+\mbf{n}_{m,l}.
\end{equation}

By applying the detection vector $\mbf{v}_{m,l}$ on the received signal, we can easily obtain
\begin{equation} \label{eq1}
{\mbf{v}_{m,l}^H}\mbf{y}_{m,l}={\mbf{v}_{m,l}^H}\mbf{H}_{m,l}\mbf{p}_m \sum_{l=1}^{L}\sqrt{\Omega_{m,l}}s_{m,l}
  + \ub{ \sum_{k=1,k \neq m}^{M} {\mbf{v}_{m,l}^H}\mbf{H}_{m,l}\mbf{p}_k \mbf{s}_k}_{\text{interference from other clusters}}+ {\mbf{v}_{m,l}^H}\mbf{n}_{m,l},
\end{equation}
where $\mbf{s}_k$ denotes the $k$th row of $\mbf{s}$.

Due to the constraint\footnote[1]{\textcolor{black}{Owing to the specific selection of $\mathbf{P}$, this constraint is further reduced to $\mathbf{v}^H_{m,l}\tilde{\mathbf{H}}_{m,l}=0$, where $\tilde{\mathbf{H}}_{m,l} = [\mathbf{h}_{1,ml}\cdots \mathbf{h}_{m-1,ml}\,\mathbf{h}_{m+1,ml}\cdots \mathbf{h}_{M,ml}]$  and $\mathbf{h}_{i,ml}$ is the $i$th column of $\mathbf{H}_{m,l}$ \cite{6}. Hence, $\mathbf{v}_{m,l}$ can be expressed as $\mathbf{U}_{m,l}\mathbf{w}_{m,l}$, where $\mathbf{U}_{m,l}$ is the matrix consisting of the left singular vectors of $\tilde{\mathbf{H}}_{m,l}$ corresponding to the non-zero singular values, and $\mathbf{w}_{m,l}$ is the maximum ratio combining vector expressed as $\mathbf{U}^H_{m,l}\mathbf{h}_{m,ml}/|\mathbf{U}^H_{m,l}\mathbf{h}_{m,ml}|$.}} on the detection vector, i.e., ${\mbf{v}_{m,l}^H}\mbf{H}_{m,l}\mbf{p}_k=0$ for any $k\neq m$, the above equation can be simplified as
%\begin{equation}
%{\mbf{v}_{m,l}^H}\mbf{y}_{m,l}={\mbf{v}_{m,l}^H}\mbf{H}_{m,l}\mbf{p}_m (\Omega_{m,1}s_{m,1}+\dots + \sqrt{\Omega_{m,l}}s_{m,L})
%  + {\mbf{v}_{m,l}^H}\mbf{n}_{m,l}
%\end{equation}
\begin{equation}
{\mbf{v}_{m,l}^H}\mbf{y}_{m,l}={\mbf{v}_{m,l}^H}\mbf{H}_{m,l}\mbf{p}_m \sum_{l=1}^{L}\sqrt{\Omega_{m,l}}s_{m,l}
  + {\mbf{v}_{m,l}^H}\mbf{n}_{m,l}.
\end{equation}

Without loss of generality, the effective channel gains are rearranged as
\begin{equation} \label{eq:order}
|{\mbf{v}_{m,1}^H}\mbf{H}_{m,1}\mbf{p}_m|^2 \geq \dots \geq |{\mbf{v}_{m,L}^H}\mbf{H}_{m,L}\mbf{p}_m|^2.
\end{equation}

At the receiver side, SIC will be conducted by user $(m,l)$ to remove the interference from the users with worse channel gains, i.e., $(m,l+1), \dots, (m,L)$. At this juncture, the following lemma is helpful to understand the efficient performance of SIC at user $(m,l)$.

\begin{lemma}
The interference from user $(m,k), \forall{k}\in \{l+1,\dots,L \}$ can be removed at user $(m,l)$.  
\end{lemma}

\begin{IEEEproof}
Refer to Appendix A.
\end{IEEEproof}

\begin{remark}
Lemma 1 shows that under the given system model, the interference from users with worse channel conditions can be removed. Consequently, the achieved data rate at user $(m,l)$ is given by
\begin{equation} \label{eq:NOMA_l}
R_{m,l}^{\text{NOMA}}=\log_2
\begin{pmatrix}
1+ \frac{\rho \Omega_{m,l} |{\mbf{v}_{m,l}^H}\mbf{H}_{m,l}\mbf{p}_m|^2 }
{1+  \rho \sum_{k=1}^{l-1} \Omega_{m,k} |{\mbf{v}_{m,l}^H}\mbf{H}_{m,l}\mbf{p}_m|^2}
\end{pmatrix},
\end{equation}
where {\color{black}$\rho=1/\sigma_n^2$, with $\sigma_n^2$ as the noise variance. We assume that the noise variance is the same for all users.}
%where $\rho=\frac{1}{|\mbf{v}_{m,l}^H \mbf{n}_{m,l}|^2}$ is the same for all users, since the detection vector is normalized and the noise variance remains unchanged after rotation. 
\end{remark}

\subsection{\ac{MIMO}-\ac{OMA}}
{\color{black}For the \ac{OMA} scheme, the same power coefficients are allocated to the $L$ users per cluster as for the case of \ac{MIMO}-\ac{NOMA} for the sake of comparison, i.e., $\Omega_{m,1}, \dots, \Omega_{m,L}$. In addition, the degrees of freedom (time or frequency) are split amongst the $L$ users per cluster, i.e., user $(m,l)$ is assigned a fraction of the degrees of freedom, denoted by $\lambda_{m,l}$, satisfying $\sum_{l=1}^{L}\lambda_{m,l}=1$. Accordingly, the achieved data rate at user $(m,l)$ is given by \cite{4}
}

%For the \ac{OMA} scheme, the degrees of freedom (time or frequency) are split amongst the $L$ users per cluster. For the sake of comparison, the same power coefficients are allocated to the $L$ users per cluster as for the case of \ac{MIMO}-\ac{NOMA}, i.e., $\Omega_{m,1}, \dots, \Omega_{m,L}$. In addition, user $(m,l)$ is assigned a fraction of the degrees of freedom, denoted by $\lambda_{m,l}$, satisfying $\sum_{l=1}^{L}\lambda_{m,l}=1$. Accordingly, the achieved data rate at user $(m,l)$ is given by \cite{4}

\begin{equation}
R_{m,l}^{\text{OMA}}=\lambda_{m,l} \log_2
\begin{pmatrix}
1+ \frac{\rho \Omega_{m,l} |{\mbf{v}_{m,l}^H}\mbf{H}_{m,l}\mbf{p}_m|^2 }
{\lambda_{m,l}}
\end{pmatrix}. 
\end{equation}

The following lemma gives the {\color{black}sum rate} upper bound when two users are paired in a cluster. 

\begin{lemma}
%The sum rate for two users, denoted as $C_{m,2}^{\text{OMA}}$ is bounded by
The {\color{black}sum rate} for two users ${\color{black}S_{m,2}^{\text{OMA}} }$ is bounded \text{by \cite{18}}
\begin{eqnarray} \label{eq:2OMA}
{\color{black}S_{m,2}^{\text{OMA}} } \leq \log_2 (1+  \sum_{l=1}^{2} \rho \Omega_{m,l} |{\mbf{v}_{m,l}^H}\mbf{H}_{m,l}\mbf{p}_m|^2  ),
\end{eqnarray}
where the equality holds when
\begin{equation} \label{eq:2OMA_cond}
{\lambda_{m,l}=\frac{ \Omega_{m,l} |{\mbf{v}_{m,l}^H}\mbf{H}_{m,l}\mbf{p}_m|^2 } { \sum_{k=1}^{2}  \Omega_{m,k} |{\mbf{v}_{m,k}^H}\mbf{H}_{m,k}\mbf{p}_m|^2} }, l \in \{1,2\}.
\end{equation}
\end{lemma}

%\begin{IEEEproof}
%Refer to Appendix B.
%\end{IEEEproof}

%\begin{IEEEproof}
%The sum rate for the two users is given by
%\begin{multline} \label{eq:lemma2_prof}
%C_{m,2}^{\text{OMA}}=\lambda_{m,1} \log_2
%\begin{pmatrix}
%1+ \frac{\rho \Omega_{m,1} |{\mbf{v}_{m,1}^H}\mbf{H}_{m,1}\mbf{p}_m|^2 }
%{\lambda_{m,1}}
%\end{pmatrix} \\
%+
%(1-\lambda_{m,1}) \log_2
%\begin{pmatrix}
%1+ \frac{\rho \Omega_{m,2} |{\mbf{v}_{m,2}^H}\mbf{H}_{m,2}\mbf{p}_m|^2 }
%{(1-\lambda_{m,1})}
%\end{pmatrix}.
%\end{multline}
%
%It is shown in \cite{18} that when (\ref{eq:2OMA_cond}) is satisfied, $C_{m,2}^{\text{OMA}}$ is maximized, and (\ref{eq:2OMA}) is obtained. Thus, Lemma 2 is proved.
%
%%After some simple mathematical manipulations, it can be easily obtained that when (\ref{eq:2OMA_cond}) is satisfied, $C_{m,2}^{\text{OMA}}$ is maximized, and (\ref{eq:2OMA}) is obtained. Thus, Lemma 2 is proved.
%\end{IEEEproof}

\begin{remark}
Lemma 2 gives the maximum {\color{black}sum rate} of two users for MIMO-OMA. On this basis, the bound of the {\color{black}sum rate} for the $m$th cluster can be derived, when there are $L$ users. 
\end{remark}

\begin{theorem}
%The total capacity in the $m$th cluster, denoted as $C_{m}^{\text{OMA}}=\sum_{l=1}^{L}R_{m,l}^{\text{OMA}}$, is bounded by
The {\color{black}sum rate} in the $m$th cluster is upper bounded by
\begin{equation} \label{eq:LOMA}
{\color{black}S_{m,L}^{\text{OMA}} } \leq \log_2 (1+ \sum_{l=1}^{L}  \rho \Omega_{m,l} |{\mbf{v}_{m,l}^H}\mbf{H}_{m,l}\mbf{p}_m|^2 ),
\end{equation}
where the equality holds when 
\begin{equation}\label{eq:LOMA_cond}
{\lambda_{m,l}=\frac{ \Omega_{m,l} |{\mbf{v}_{m,l}^H}\mbf{H}_{m,l}\mbf{p}_m|^2 } { \sum_{k=1}^{L}  \Omega_{m,k} |{\mbf{v}_{m,k}^H}\mbf{H}_{m,k}\mbf{p}_m|^2} }, l\in\{1,\dots,L\}.
\end{equation}

\end{theorem}

\begin{IEEEproof}
Refer to Appendix B.
\end{IEEEproof}

\begin{remark}
Theorem 1 shows that once the power allocation coefficients are ascertained, the optimal allocation of degrees of freedom can be obtained accordingly to ensure that the maximum {\color{black}sum rate} for the $m$th cluster ${\color{black}S_{m,L}^{\text{OMA}} }$ is achieved. 
\end{remark}

\section{Capacity Comparison between \ac{MIMO}-\ac{NOMA} and \ac{MIMO}-\ac{OMA}}
In this section, both {\color{black}sum channel capacity} and {\color{black}ergodic sum capacity} for the $m$th cluster achieved by \ac{MIMO}-\ac{NOMA} are compared to that achieved by \ac{MIMO}-\ac{OMA}. 

%It is proved that \ac{MIMO}-\ac{NOMA} strictly outperforms \ac{MIMO}-\ac{OMA} in terms of {\color{black}sum rate}, i.e., for any power split in \ac{MIMO}-\ac{OMA}, a higher capacity can be obtained by \ac{MIMO}-\ac{NOMA} via allocating the same power split to the latter. On this basis, it is straightforward to conclude that a higher {\color{black}ergodic sum rate} can also be achieved by \ac{MIMO}-\ac{NOMA} when compared with \ac{MIMO}-\ac{OMA}. 

\subsection{{\color{black}Sum Channel Capacity}}
The {\color{black}sum rate} for \ac{MIMO}-\ac{OMA} has already been obtained, i.e., (\ref{eq:LOMA}) and (\ref{eq:LOMA_cond}). Now, the {\color{black}sum rate} for the $m$th cluster in \ac{MIMO}-\ac{NOMA} is considered, which is ${\color{black}S_{m,L}^{\text{NOMA}} }=\sum_{l=1}^{L}R_{m,l}^{\text{NOMA}}$, and can be easily expressed as
%\begin{equation}
%\begin{split}
% &C_{m}^{\text{NOMA}} \\
%&= \sum_{l=1}^{L} \log_2
%\begin{pmatrix}
%1+ \frac{\rho \Omega_{m,l} |{\mbf{v}_{m,l}^H}\mbf{H}_{m,l}\mbf{p}_m|^2 }
%{1+  \rho \sum_{k=1}^{l-l} \Omega_{m,k} |{\mbf{v}_{m,k}^H}\mbf{H}_{m,k}\mbf{p}_m|^2}
%\end{pmatrix}
%\end{split}
%\end{equation}
\begin{equation} \label{eq:NOMA_sum}
{\color{black}S_{m,L}^{\text{NOMA}} } = \sum_{l=1}^{L} \log_2
\begin{pmatrix}
1+ \frac{\rho \Omega_{m,l} |{\mbf{v}_{m,l}^H}\mbf{H}_{m,l}\mbf{p}_m|^2 }
{1+  \rho \sum_{k=1}^{l-l} \Omega_{m,k} |{\mbf{v}_{m,l}^H}\mbf{H}_{m,l}\mbf{p}_m|^2}
\end{pmatrix}.
\end{equation}

%\end{multline}

\begin{lemma}
The lower bound of the {\color{black}sum rate} for \ac{MIMO}-\ac{NOMA} is given by
\begin{equation} \label{eq:NOMA}
{\color{black}S_{m,L}^{\text{NOMA}} } \geq \log_2(1+ \rho \sum_{l=1}^{L}  \Omega_{m,l} |{\mbf{v}_{m,l}^H}\mbf{H}_{m,l}\mbf{p}_m|^2  ).
\end{equation}
\end{lemma}

\begin{IEEEproof}
Refer to Appendix C.
\end{IEEEproof}

\begin{theorem}
For any power split in \ac{MIMO}-\ac{OMA}, a larger {\color{black}sum rate can be achieved by \ac{MIMO}-\ac{NOMA} via assigning the same power split to the latter. In particular, when the power split is optimal for MIMO-OMA, a larger sum channel capacity can be achieved by \ac{MIMO}-\ac{NOMA}.
}  
\end{theorem}

\begin{IEEEproof}
Combining Theorem 1 and Lemma 3, i.e., (\ref{eq:LOMA}) and (\ref{eq:NOMA}), we obtain 
\begin{equation} \label{eq:sum rate comp}
{\color{black}S_{m,L}^{\text{NOMA}} \geq  S_{m,L}^{\text{OMA}},} 
\end{equation}
which proves the superiority of \ac{MIMO}-\ac{NOMA} over \ac{MIMO}-\ac{OMA} in terms of sum rate for any power split. 

When the power split is optimal for MIMO-OMA, the sum channel capacity, denoted as $C_{m,L}^{\text{OMA}}$, is achieved if (\ref{eq:LOMA_cond}) is met. Let us assign the same power split to MIMO-NOMA and denote its sum rate as ${S}_{m,L}^{'\text{NOMA}}$. We also denote the sum channel capacity for MIMO-NOMA as $C_{m,L}^{\text{NOMA}}$, which satisfies $C_{m,L}^{\text{NOMA}} \geq {S}_{m,L}^{'\text{NOMA}}$. Thus, we have
\begin{equation}
{\color{black}C_{m,L}^{\text{NOMA}} \geq {S}_{m,L}^{'\text{NOMA}} \geq C_{m,L}^{\text{OMA}},} 
\end{equation}
where the second inequality comes from (\ref{eq:sum rate comp}). Therefore, MIMO-NOMA achieves a larger sum channel capacity than MIMO-OMA.
\end{IEEEproof}

In summary, it is proved analytically that for any instantaneous channel gain $\mbf{H}_{m,l} ~ (m \in \{1,\dots,M\}, l \in \{1,\dots,L\})$, given the power split in \ac{MIMO}-\ac{OMA}, a larger {\color{black}sum rate} can be achieved by \ac{MIMO}-\ac{NOMA} via simply allocating the same power split to the latter. Note that there is no constraint on the value of the power split, which means that the conclusion is true for any power split. Therefore, we can conclude that even when there are multiple users per cluster, \ac{MIMO}-\ac{NOMA} strictly outperforms \ac{MIMO}-\ac{OMA} in terms of the {\color{black}sum rate} under any instantaneous channel gain $\mbf{H}_{m,l}$ and any power split. {\color{black}On this basis, it is shown that MIMO-NOMA also achieves a larger sum channel capacity than MIMO-OMA.
}

Furthermore, when there are only two users per cluster, the following lemma provides the power allocation coefficient such that the gap between the {\color{black}sum rate} of \ac{MIMO}-\ac{NOMA} and \ac{MIMO}-\ac{OMA} is maximized.
\begin{lemma}
The {\color{black}sum rate} gap for two users between \ac{MIMO}-\ac{NOMA} and \ac{MIMO}-\ac{OMA} is maximized, when the following equation is satisfied
\begin{equation} \label{eq:gap_cond}
\Omega_{m,1}=\frac{ \sqrt{\rho|{\mbf{v}_{m,1}^H}\mbf{H}_{m,1}\mbf{p}_m|^2+1}-1}{\rho|{\mbf{v}_{m,1}^H}\mbf{H}_{m,1}\mbf{p}_m|^2}.
\end{equation}
\end{lemma}

\begin{IEEEproof}
According to (\ref{eq:2OMA}) and (\ref{eq:NOMA_sum}), the sum rate gap between \ac{MIMO}-\ac{NOMA} and \ac{MIMO}-\ac{OMA} is given by
\begin{equation} \label{eq:gap}
\begin{split}
\bigtriangleup S_{m,2}=&\log_2 \{1+\rho \Omega_{m,1} |{\mbf{v}_{m,1}^H}\mbf{H}_{m,1}\mbf{p}_m|^2 \} \\
&+\log_2 \bigg\{1+\frac{\rho \Omega_{m,2} |{\mbf{v}_{m,2}^H}\mbf{H}_{m,2}\mbf{p}_m|^2 }{1+\rho \Omega_{m,1} |{\mbf{v}_{m,2}^H}\mbf{H}_{m,2}\mbf{p}_m|^2} \bigg\} \\
&-  \log_2 (1+  \sum_{l=1}^{2} \rho \Omega_{m,l} |{\mbf{v}_{m,l}^H}\mbf{H}_{m,l}\mbf{p}_m|^2  ).
\end{split}
\end{equation}

After replacing $\Omega_{m,2}$ with $1-\Omega_{m,1}$, the only variable is $\Omega_{m,1}$. It can be easily proved that when (\ref{eq:gap_cond}) is satisfied, $\frac{\partial\bigtriangleup S_{m,2} }{\partial \Omega_{m,1}}=0$. Moreover, $\frac{\partial\bigtriangleup S_{m,2} }{\partial \Omega_{m,1}}>0$ when $\Omega_{m,1} < \frac{ \sqrt{\rho|{\mbf{v}_{m,1}^H}\mbf{H}_{m,1}\mbf{p}_m|^2+1}-1}{\rho|{\mbf{v}_{m,1}^H}\mbf{H}_{m,1}\mbf{p}_m|^2}$, and $\frac{\partial\bigtriangleup S_{m,2} }{\partial \Omega_{m,1}}<0$, otherwise. Therefore, the sum rate gap is maximized when (\ref{eq:gap_cond}) holds. In addition, since $\rho|{\mbf{v}_{m,1}^H}\mbf{H}_{m,1}\mbf{p}_m|^2>0$, it can be easily proven that $0<\frac{ \sqrt{\rho|{\mbf{v}_{m,1}^H}\mbf{H}_{m,1}\mbf{p}_m|^2+1}-1}{\rho|{\mbf{v}_{m,1}^H}\mbf{H}_{m,1}\mbf{p}_m|^2}<1$, which fits the range of $\Omega_{m,1}$. 
\end{IEEEproof}

{\color{black}Accordingly, for the two user case, we can calculate the maximum {\color{black}sum rate} gap between MIMO-NOMA and MIMO-OMA by substituting the value of $\Omega_{m,1}$ from (\ref{eq:gap_cond}) into (\ref{eq:gap}).}

\begin{remark}
It is somewhat surprising that the power coefficient maximizing the {\color{black}sum rate} gap is only determined by the channel of the first user. Moreover, according to (\ref{eq:gap_cond}), it can be easily verified that $\Omega_{m,1}$ declines with $\rho|{\mbf{v}_{m,1}^H}\mbf{H}_{m,1}\mbf{p}_m|^2$. Specifically, when $\rho|{\mbf{v}_{m,1}^H}\mbf{H}_{m,1}\mbf{p}_m|^2\rightarrow 0$, $\Omega_{m,1}\rightarrow 0.5$, and $\rho|{\mbf{v}_{m,1}^H}\mbf{H}_{m,1}\mbf{p}_m|^2\rightarrow \infty$, $\Omega_{m,1}\rightarrow 0$. Thus, it can be further concluded that $\Omega_{m,1}<0.5$ for any value of $\rho|{\mbf{v}_{m,1}^H}\mbf{H}_{m,1}\mbf{p}_m|^2$. This is consistent with the concept of NOMA, in which a larger proportion of power should be allocated to the user with worse channel condition.
\end{remark}

\subsection{{\color{black}Ergodic Sum Capacity}}
\begin{corollary}
For any power split in \ac{MIMO}-\ac{OMA}, a larger {\color{black}ergodic sum rate} can be achieved by \ac{MIMO}-\ac{NOMA} via assigning the same power split to the latter. {\color{black}In particular, when the power split is optimal for MIMO-OMA, a larger ergodic sum capacity can be achieved by \ac{MIMO}-\ac{NOMA}.}
\end{corollary}

\begin{IEEEproof}
As shown in the previous section, \ac{MIMO}-\ac{NOMA} strictly outperforms \ac{MIMO}-\ac{OMA} in terms of {\color{black}sum rate} under any instantaneous channel gains of $\mbf{H}_{m,l}$. By applying the expectation operator, it is straightforward to claim that the {\color{black}ergodic sum rate} of \ac{MIMO}-\ac{NOMA} is always larger than that of \ac{MIMO}-\ac{OMA}. {\color{black}Likewise, it is easy to verify that the {\color{black}ergodic sum capacity} of \ac{MIMO}-\ac{NOMA} is always larger than that of \ac{MIMO}-\ac{OMA}.} Additionally, it is worth noticing that the conclusions hold regardless of the distribution of $\mbf{H}_{m,l}$.
\end{IEEEproof}  

To summarize, the same conclusion as for the {\color{black}sum channel capacity} holds true for the {\color{black}ergodic sum capacity}. Thus, even for the case of multiple users per cluster, \ac{MIMO}-\ac{NOMA} strictly outperforms \ac{MIMO}-\ac{OMA} in terms of both {\color{black}sum channel capacity} and {\color{black}ergodic sum capacity}. 
  
\section{User Admission}  
Analytical results obtained in the previous section validate that \ac{MIMO}-\ac{NOMA} strictly outperforms \ac{MIMO}-\ac{OMA} in terms of both {\color{black}sum rate} and {\color{black}ergodic sum rate}, even when there are multiple users in a cluster. Does this mean we should group a large number of users in a cluster to increase the system capacity in terms of the number of users? Clearly, \ac{SIC} at the receiver becomes increasingly complicated when more users are included in a cluster, which limits the practical number of users per cluster. Furthermore, the study of how the {\color{black}sum rate} varies with the number of admitted users is of interest, which we explore in the following section.

%It is validated in the above section that \ac{MIMO}-\ac{NOMA} strictly outperforms \ac{MIMO}-\ac{OMA} in terms of both {\color{black}sum rate} and {\color{black}ergodic sum rate}, even when there are multiple users in a cluster. Does this mean we should group a large number of users in a cluster to increase the system capacity in terms of the number of users? Firstly, \ac{SIC} at the receiver side of the users becomes increasingly complicated when more users are included in a cluster, which limits the practical number of users in a cluster. Furthermore, as will be shown in the following section, when more users are grouped into a cluster, a lower total throughput is achieved. Thus, a tradeoff between the number of admitted users and the total {\color{black}sum rate} should be considered. On this basis, a user admission scheme is proposed here. When the \ac{SINR} thresholds of the users are equal, the proposed user admission scheme is optimal in terms of both {\color{black}sum rate} and number of admitted users. Otherwise, it still achieves a good performance in balancing both criteria. Moreover, its computational complexity is only $O(N)$. 

\subsection{{\color{black}Sum Rate} versus Number of Users}
Here the MIMO-NOMA {\color{black}sum rate} between the case of $l$ and $l+1$ users in the $m$th cluster is compared. For notational simplicity, the index of the cluster, $m$, and the NOMA superscript are omitted. The power allocation coefficients for $1$-to-$l$ and $1$-to-$(l+1)$ users are denoted as $\Omega_{1},\dots,\Omega_{l}$ and $\Theta_{1},\dots,\Theta_{l+1}$ respectively, satisfying $\sum_{k=1}^{l}\Omega_{k}=\sum_{k=1}^{l+1} \Theta_{k}=1$, and $\Omega_{k} \geq \Theta_{k}, \forall k \in \{1, \dots, l\}$. Additionally, we set $\Xi_k=\rho |{\mbf{v}_{k}^H}\mbf{H}_{k}\mbf{p}|^2, k\in \{1, \dots, l+1 \}$ for notational simplicity, and the effective channel of the users follow the order in (\ref{eq:order}), i.e., $\Xi_1 \geq \dots \geq \Xi_{l+1}$.

According to (\ref{eq:NOMA_l}), the {\color{black}sum rate} up to $l$ users can be easily re-written as
\begin{equation} \label{eq:14}
\begin{split}
S^{(l)}&=\sum_{k=1}^{l}R_{k}^{(l)} \\ 
&= \log_2(1+\Omega_{1} \Xi_1)+ \sum_{k=2}^{l} \log_2 \bigg( \frac{ 1+ \sum_{i=1}^{k} \Omega_{i} \Xi_k } {1+ \sum_{i=1}^{k-1} \Omega_{i} \Xi_k }\bigg),
\end{split}
\end{equation} 
%\begin{equation} \label{eq:14}
%C^{(l)}= \log_2(1+\Omega_{1} \Xi_1)+ \sum_{k=2}^{l} \log_2 \bigg( \frac{ 1+ \sum_{i=1}^{k} \Omega_{i} \Xi_k } {1+ \sum_{i=1}^{k-1} \Omega_{i} \Xi_k }\bigg),
%\end{split}
%\end{equation} 
where $R_{k}^{(l)}$ denotes the rate of the $k$th user for the case of $l$ users in total. 

Likewise, the {\color{black}sum rate} for the $l+1$ users can be expressed as
\begin{equation} \label{eq:15}
\begin{split}
S^{(l+1)}=&\sum_{k=1}^{l+1}R_{k}^{(l+1)} \\ 
=& \log_2(1+\Theta_1 \Xi_1)+ \sum_{k=2}^{l} \log_2  \frac{1+ \sum_{i=1}^{k} \Theta_i \Xi_k } {1+ \sum_{i=1}^{k-1} \Theta_i \Xi_k } \\
&+ \log_2  \frac{1+\Xi_{l+1} } {1+ \sum_{i=1}^{l} \Theta_i \Xi_{l+1} } ,
\end{split}
\end{equation} 
%\begin{equation} \label{eq:15}
%\begin{split}
%C^{(l+1)}=& \log_2(1+\Theta_1 \Xi_1)+ \sum_{k=2}^{l} \log_2  \frac{1+ \sum_{i=1}^{k} \Theta_i \Xi_k } {1+ \sum_{i=1}^{k-1} \Theta_i \Xi_k } \\
%&+ \log_2  \frac{1+\Xi_{l+1} } {1+ \sum_{i=1}^{l} \Theta_i \Xi_{l+1} }.
%\end{split}
%\end{equation} 
where $R_{k}^{(l+1)}$ denotes the rate of the $k$th user for the case of $l+1$ users in total. 

Combining (\ref{eq:14}) and (\ref{eq:15}), the difference between the two sum rates, denoted by $\Lambda=S^{(l+1)}-S^{(l)}$, can be expressed as

\begin{equation} \label{eq:dC}
\begin{split}
\Lambda=& \log_2 \frac{1+\Theta_1 \Xi_1}{1+\Omega_{1} \Xi_1} + \log_2 \frac{1+\Xi_{l+1} } {1+ \sum_{i=1}^{l} \Theta_i \Xi_{l+1} } \\
& + \sum_{k=2}^{l} \log_2\frac{1+ \sum_{i=1}^{k} \Theta_i \Xi_k } {1+ \sum_{i=1}^{k-1} \Theta_i \Xi_k } 
 \times  \frac{1+ \sum_{i=1}^{k-1} \Omega_{i} \Xi_k }{ 1+ \sum_{i=1}^{k} \Omega_{i} \Xi_k }  \\ 
=& \log_2\frac{1+\Theta_1 \Xi_1}{1+\Omega_{1} \Xi_1} + \log_2 \frac{1+\Xi_{l+1} } {1+ \sum_{i=1}^{l} \Theta_i \Xi_{l+1} } \\
&+ \sum_{k=2}^{l} \log_2\frac{1+ \sum_{i=1}^{k} \Theta_i \Xi_k }{ 1+ \sum_{i=1}^{k} \Omega_{i} \Xi_k }  \times  \frac{1+ \sum_{i=1}^{k-1} \Omega_{i} \Xi_k } {1+ \sum_{i=1}^{k-1} \Theta_i \Xi_k }  \\ 
= & \log_2 \bigg\{ \frac{1+\Theta_1 \Xi_1}{1+\Omega_{1} \Xi_1} \times \frac{1+\Xi_{l+1} } {1+ \sum_{i=1}^{l} \Theta_i \Xi_{l+1} } \\
 &\times \prod_{k=2}^{l}  \frac{1+ \sum_{i=1}^{k} \Theta_i \Xi_k }{ 1+ \sum_{i=1}^{k} \Omega_{i} \Xi_k } \times  \frac{1+ \sum_{i=1}^{k-1} \Omega_{i} \Xi_k } {1+ \sum_{i=1}^{k-1} \Theta_i \Xi_k } 
 \bigg \} \\ 
= & \log_2 \bigg \{ \ub{ \frac{1+\Theta_1 \Xi_1}{1+\Omega_{1} \Xi_1} \times \frac{1+\Omega_{1} \Xi_2}{1+ \Theta_1 \Xi_2} }_{\Lambda_1} \\
& \times \ub{ \prod_{k=2}^{l-1}  \frac{1+ \sum_{i=1}^{k} \Theta_i \Xi_k }{ 1+ \sum_{i=1}^{k} \Omega_{i} \Xi_k }  \times  \frac{1+ \sum_{i=1}^{k} \Omega_{i} \Xi_{k+1} } {1+ \sum_{i=1}^{k} \Theta_i \Xi_{k+1} }  }_{\Lambda_2}   \\ 
& \times \ub{ \frac{1+ \sum_{i=1}^{l} \Theta_i \Xi_l }{ 1+ \sum_{i=1}^{l} \Omega_{i} \Xi_l } \times \frac{1+\Xi_{l+1} } {1+ \sum_{i=1}^{l} \Theta_i \Xi_{l+1} } }_{\Lambda_3} \bigg \}.
\end{split}
\end{equation}

First, let us consider $\Lambda_1$, which is given by
\begin{equation}
\begin{split}
 \Lambda_1= \frac{1+\Theta_1 \Xi_1+\Omega_{1} \Xi_2+\Theta_1 \Xi_1\Omega_{1} \Xi_2 }{1+\Omega_{1} \Xi_1+ \Theta_1 \Xi_2+\Omega_{1} \Xi_1\Theta_1 \Xi_2}.
\end{split}
\end{equation}

Due to $(\Xi_1-\Xi_2)(\Theta_1-\Omega_{1})\leq 0$, it can be easily shown that $\Lambda_1\leq 1$. 

Likewise, the same method for $\Lambda_2$ can be applied. Indeed, owing to $\sum_{i=1}^{k}(\Theta_i-\Omega_{i})(\Xi_k-\Xi_{k+1}) \leq 0$, it can be easily verified that each element in $\Lambda_2$ does not exceed 1. Thus, it is obtained $\Lambda_2\leq 1$.  

%Further, let us consider the elements in part 2, which are given as 
%\begin{equation}
%\begin{split}
%& \frac{1+ \sum_{i=1}^{k} \Theta_i \Xi_k }{ 1+ \sum_{i=1}^{k} \Omega_{i} \Xi_k }  \times  \frac{1+ \sum_{i=1}^{k} \Omega_{i} \Xi_{k+1} } {1+ \sum_{i=1}^{k} \Theta_i \Xi_{k+1} }   \\ 
%=& \frac{1+\sum_{i=1}^{k} \Theta_i \Xi_k+\sum_{i=1}^{k} \Omega_{i} \Xi_{k+1} +\sum_{i=1}^{k}  \Theta_i \Xi_k \times \sum_{i=1}^{k} \Omega_{i} \Xi_{k+1}}
%{1+ \sum_{i=1}^{k} \Omega_{i} \Xi_k +\sum_{i=1}^{k} \Theta_i \Xi_{k+1} + \sum_{i=1}^{k} \Omega_{i} \Xi_k \times \sum_{i=1}^{k} \Theta_i \Xi_{k+1}},
%\end{split}
%\end{equation}
%where $k\in \{2, \dots, l-1 \}$.
%
%Due to $\sum_{i=1}^{k}(\Theta_i-\Omega_{i})(\Xi_k-\Xi_{k+1}) \leq 0$, each element in part 2 does not exceed 1. Thus, part 2 does not exceed 1. 
%
%In terms of part 1, we can use the same method, and it is easy to verify that part 1 also does not exceed 1. 

As for $\Lambda_3$, by applying $\sum_{i=1}^{l} \Omega_{i}=1$, we have 
\begin{equation}
\begin{split}
\Lambda_3 = \frac{1+\sum_{i=1}^{l} \Theta_i \Xi_l+\Xi_{l+1}+ \sum_{i=1}^{l} \Theta_i \Xi_l \Xi_{l+1} }{1+\sum_{i=1}^{l} \Theta_i \Xi_{l+1}+ \Xi_l +\sum_{i=1}^{l} \Theta_i \Xi_l \Xi_{l+1} }.
\end{split}
\end{equation}

As $(\Xi_l-\Xi_{l+1})(\sum_{i=1}^{l} \Theta_i-1) \leq 0$, then $\Lambda_3 \leq 1$. By combining the results for $\Lambda_1, \Lambda_2$ and $\Lambda_3$ in (\ref{eq:dC}), it leads to $\Lambda \leq 0$. 

To conclude, the more users are admitted, the lower the {\color{black}sum rate} is obtained. This requires further consideration of the tradeoff between the {\color{black}sum rate} and number of admitted users. We will thus consider the problem of maximizing the user admission when the users \ac{SINR} thresholds are given.  
 
\subsection{Proposed User Admission Scheme} 
The \ac{SINR} thresholds of the $L$ users in the $m$th cluster are denoted as $\Gamma_{1},\dots, \Gamma_{L}$. In addition, the maximum number of admitted users is represented as $l, l\in \{0, 1, \dots, L \}$. Further, the $l$ admitted users are denoted as $a_1, a_2, \dots, a_{l}$. Accordingly, the problem can be formulated as 
%\begin{subequations} \label{eq:user}
%\begin{align}
% &{}^{\displaystyle{\max  \: l }}_{\mbf{\Omega}}  \\
%\text{subject to}\nonumber \\
%    & \gamma_k \geq \Gamma_k,  k\in \{ a_1, a_2, \dots, a_{l} \} \\
%   & \sum_{k=a_1}^{a_{l}}{\Omega_k} \leq 1,
%\end{align}
%\end{subequations}
\begin{IEEEeqnarray*}{clr}\label{eq:user}
\displaystyle\underset{\Omega}{\text{max}}  & \quad l \IEEEyesnumber \IEEEyessubnumber* \\
\text{s.t.}& \quad  \gamma_k \geq \Gamma_k, &  k\in \{a_1, a_2, \ldots, a_l\} \\
& \quad \sum\limits_{k=a_1}^{a_l}\Omega_k \leq 1,
\end{IEEEeqnarray*} 
%\begin{subequations} \label{eq:user}
%\begin{align}
% \displaystyle{\max} ~_{\{\mbf{\Omega}\}} ~~ &l \\
%\text{s.t.} ~~    & \gamma_k \geq \Gamma_k,  k\in \{ a_1, a_2, \dots, a_{l} \} \\
%   & \sum_{k=a_1}^{a_{l}}{\Omega_k} \leq 1,
%\end{align}
%\end{subequations}  
where $\mbf{ \Omega }=[\Omega_{1}, \ldots, \Omega_{L}]$ is the vector whose elements are the power allocation coefficients, and $\gamma_k$ is the SINR of the $k$th admitted user, given by
\begin{equation} \label{eq:SINR}
\gamma_k=\frac{\rho \Omega_k |{\mbf{v}_{k}^H}\mbf{H}_{k}\mbf{p}|^2 }
{1+  \rho \sum_{i=1}^{k-1} \Omega_i |{\mbf{v}_{k}^H}\mbf{H}_{k}\mbf{p}|^2}.
\end{equation}

By combining (\ref{eq:user}b) and (\ref{eq:SINR}), we have
%\begin{subequations}
\begin{align} \label{eq:PA}
   \Omega_k &\geqslant \Gamma_k \sum_{i=1}^{k-1}{\Omega_i} +\frac{\Gamma_k}{\rho |{\mbf{v}_{k}^H}\mbf{H}_{k}\mbf{p}|^2 },  
\end{align}
%\end{subequations}
where variables are only $\sum_{i=1}^{k-1}{\Omega_i}$, since the other parameters, i.e., $\rho$, $\Gamma_k$, and $|{\mbf{v}_{k}^H}\mbf{H}_{k}\mbf{p}|^2 $, are known at the BS. Therefore, if the power coefficient among users is allocated in an ascending order, i.e., from the $1$st user to the $L$th user sequentially, we can obtain the power coefficient for the $k$th user easily, since $\sum_{i=1}^{k-1}{\Omega_i}$ is already known. Specifically, the power coefficient for the $1$st user is calculated as

\begin{equation}\label{eq:24}
  \Omega_1=\frac{\Gamma_1}{\rho |{\mbf{v}_{1}^H}\mbf{H}_{1}\mbf{p}|^2 }.
\end{equation}

Sequentially and iteratively, when the power coefficient of the $1$st user is known, it is employed to allocate the power coefficient to the $2$nd user. According to (\ref{eq:PA}), we have
\begin{equation}\label{eq:25}
 \Omega_2=\Gamma_2\Omega_1+\frac{\Gamma_2}{\rho |{\mbf{v}_{2}^H}\mbf{H}_{2}\mbf{p}|^2 }.
\end{equation}

Likewise, the power coefficient for the $k$th  user can be expressed as
\begin{equation}\label{eq:PA2}
   \Omega_k=\Gamma_k \sum_{i=1}^{k-1}{\Omega_i} +\frac{\Gamma_k}{\rho |{\mbf{v}_{k}^H}\mbf{H}_{k}\mbf{p}|^2 }.
\end{equation}

Obviously, power allocation for all users can be obtained according to (\ref{eq:PA2}). However, it should be noted that the total power constraint has not been considered yet during the user admission process above. Thus, when calculating the power coefficient for the $k$th user, we also need to ensure that the total power assigned to users, $\sum_{i=1}^{k}{\Omega_i}$, does not exceed $1$. This is obtained by comparing $\Gamma_k \sum_{i=1}^{k-1}{\Omega_i} +\frac{\Gamma_k}{\rho |{\mbf{v}_{k}^H}\mbf{H}_{k}\mbf{p}|^2 }$ with $1-\sum_{i=1}^{k-1}{\Omega_i}$ during each allocation phase. Whenever $\Gamma_k \sum_{i=1}^{k-1}{\Omega_k} +\frac{\Gamma_k}{\rho |{\mbf{v}_{k}^H}\mbf{H}_{k}\mbf{p}|^2 } < \: 1-\sum_{i=1}^{k-1}{\Omega_i}$, it implies that there is not enough power left to be assigned to the $k$th user to satisfy its SINR requirement. Therefore, the user admission process stops and the allocated power for the $k$th user is zero. Evidently, the same holds for $\{k+1,\dots,L\}$ users, i.e., $\Omega_i=0,i\in \{k,\dots,L\}$. The admitted users are $1$st user, $2$nd user, \dots, $(k-1)$th user, with the allocated power coefficient given by (\ref{eq:PA2}). 

As for the optimality of the proposed user admission scheme, the following theorem and corollary provide the results.

%{\color{black}{
\begin{theorem}
The proposed scheme maximizes the number of admitted users when the SINR thresholds of the users satisfy the following conditions:
\begin{subequations} \label{eq:SINRvary}
\begin{align}
&\frac{\Gamma_1}{ |{\mbf{v}_{1}^H}\mbf{H}_{1}\mbf{p}|^2 } \leq \dots \leq \frac{\Gamma_l}{ |{\mbf{v}_{l}^H}\mbf{H}_{l}\mbf{p}|^2 }  \\
&\Gamma_m \leq \Gamma_{n}, \forall m \in \{1, \dots, l\}, n \in \{l+1, \dots, L\},
\end{align}
\end{subequations}
where $l$ represents the total number of admitted users under the proposed scheme. 
\end{theorem}

\begin{IEEEproof}
Refer to Appendix D.
\end{IEEEproof}

\begin{corollary}
The proposed user admission scheme is optimal in terms of both {\color{black}sum rate} and number of admitted users when the \ac{SINR} thresholds of the users are equal.  
\end{corollary}

\begin{IEEEproof}
According to the channel ordering, namely (\ref{eq:order}), it is easy to verify that $\Gamma_k=\Gamma, k \in \{1,\dots,L\} $ satisfies both (\ref{eq:SINRvary}a) and (\ref{eq:SINRvary}b). Thus, one can conclude that the proposed user admission scheme is optimal in terms of the number of admitted users based on Theorem 3. In addition, since the \ac{SINR} thresholds of the users are equal, maximizing the number of admitted users also leads to the maximization of the {\color{black}sum rate}.   
\end{IEEEproof}

\begin{remark}
%The effectiveness of the proposed scheme is shown in the above theorem and corollary. When the \ac{SINR} thresholds of the users are equal, the proposed scheme is optimal in terms of both {\color{black}sum rate} and number of admitted users. 
When the SINR thresholds of the users are different, the proposed scheme still achieves good performance in balancing the tradeoff between {\color{black}sum rate} and number of admitted users. Specifically, when (\ref{eq:SINRvary}a) and (\ref{eq:SINRvary}b) are met, the proposed scheme maximizes the number of admitted users, although the {\color{black}sum rate} may be suboptimal. On the other hand, when (\ref{eq:SINRvary}a) is met, but (\ref{eq:SINRvary}b) is violated, namely, the \ac{SINR} thresholds of the admitted users are higher than that of the remaining users, the proposed scheme may be suboptimal in terms of the number of admitted users, while the {\color{black}sum rate} is still high due to two reasons: a) the admitted users have higher \ac{SINR} thresholds; b) as less users are admitted, less interference among users is introduced; therefore, an increased {\color{black}sum rate} is obtained.  
\end{remark}
%}}

In addition, the computational complexity of the proposed user admission scheme is only linear to the number of users per cluster.
\begin{IEEEproof}
For the proposed scheme, the user admission is carried out sequentially from the $1$st user to the $L$th user, and for each user admission process, a constant term of operations, i.e., $O(1)$,\footnote[2]{For the $k$th user, the calculation of $\sum_{i=1}^{k-1}{\Omega_i}$ seems to require $k-1$ operations. However, if we set $S_p=\sum_{i=1}^{k-1}{\Omega_i}$, $S_p$ can be updated through $S_p=S_p+\Omega_k$, and only one operation is needed. Thus, according to (\ref{eq:PA2}), only 5 operations ($2 \: '+'$, $2 \: '\times'$, and $1 \: '/'$) are needed to obtain $\Omega_k$. } is required. In all, the computational complexity is only linear to the number of users per cluster, i.e., $O(L)$.
\end{IEEEproof}

\begin{table} [!h]
\caption{Simulation Parameters.} 
\renewcommand{\arraystretch}{0.75}
\label{Table 1} 
\centering
 \begin{tabular}{c|c} 
 \hline  
\bfseries Parameters & \bfseries Value \\ [0.5ex] 
 \hline\hline
 Number of antennas &$M=3,N=3$ \\
 \hline
 Channel bandwidth & $10$ [MHz] \\ 
 \hline
Thermal noise density & $-174$ [dBm] \\
 \hline
 Path-loss model & $114+38\log_{10}(d)$, $d$ in kilometer \\
 \hline  
\end{tabular}
\end{table}

\section{Numerical Results}
In this section, simulation results are presented to verify the performance of \ac{MIMO}-\ac{NOMA} over \ac{MIMO}-\ac{OMA}, and validate the accuracy of the developed theoretical results. The parameters used in the simulations are listed in Table 
\uppercase\expandafter{\romannumeral 1}.

\begin{figure}
\centering
\includegraphics[width=0.8\textwidth]{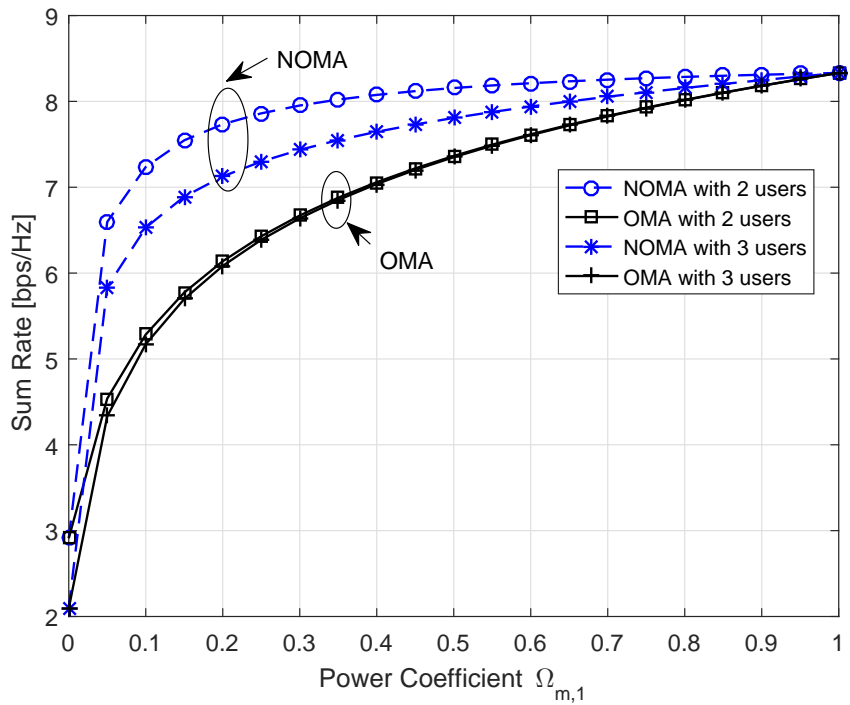}
\caption{{\color{black}Sum rate} achieved by \ac{MIMO-NOMA} and \acs{MIMO-OMA} as the power coefficient varies.}
\end{figure}
\begin{figure}
\centering
\includegraphics[width=0.8\textwidth]{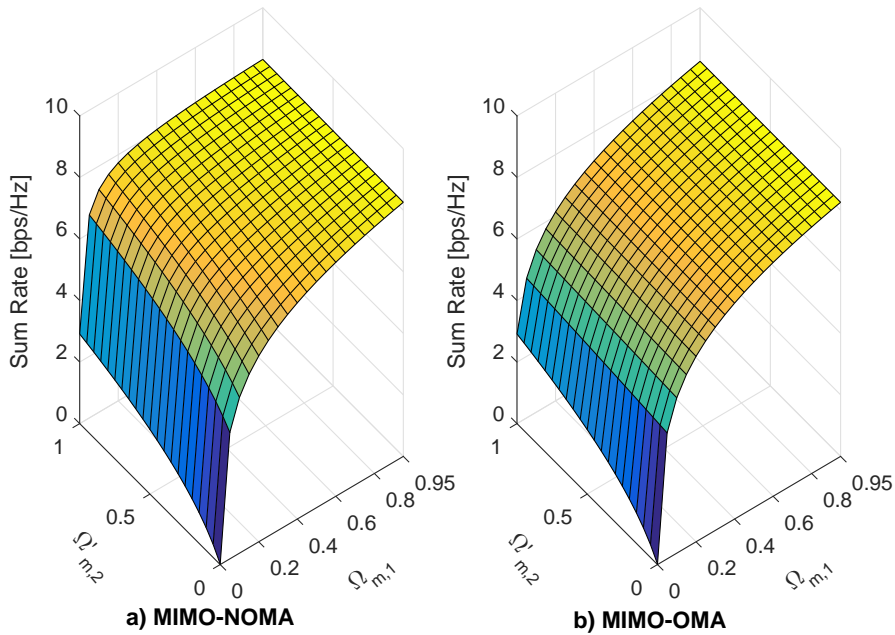}
\caption{{\color{black}Sum rate} achieved by: a) \ac{MIMO-NOMA}; b) \acs{MIMO-OMA} for 3 users as the power coefficients vary.}
\end{figure}

Fig. 1 compares the {\color{black}sum rate} of \ac{MIMO}-\ac{NOMA} and \ac{MIMO}-\ac{OMA} in two cases: with two users and three users per cluster, respectively. The total power is set to \text{35 dBm} in simulations, and $\Omega_{m,1}$ denotes the power coefficient for the first user. For the case of two users, the remaining power is allocated to the second user. For three users, the remaining power is equally divided between the second and third user. Note that the scenario that the remaining power is arbitrarily divided between the second and third user is shown in \text{Fig. 2}. Clearly, the {\color{black}sum rate} of both \ac{MIMO}-\ac{NOMA} and \ac{MIMO}-\ac{OMA} in two cases increases with $\Omega_{m,1}$, which is due to the fact that more power is allocated to the user with better channel gain. Specifically, when $\Omega_{m,1}=0$, for the two user case, the same {\color{black}sum rate} is achieved for both \ac{MIMO}-\ac{NOMA} and \ac{MIMO}-\ac{OMA}, since only the second user is being served. On the other hand, for the three users case, \ac{MIMO}-\ac{NOMA} is slightly larger than \ac{MIMO}-\ac{OMA}, since two users are being served.  In contrast, when $\Omega_{m,1}=1$, the {\color{black}sum rate} of both \ac{MIMO}-\ac{NOMA} and \ac{MIMO}-\ac{OMA} in two cases is the same since only the first user is served. In addition, for any other power split, \ac{MIMO}-\ac{NOMA} outperforms \ac{MIMO}-\ac{OMA} for both cases, which coincides with our result that \ac{MIMO}-\ac{NOMA} always has a larger {\color{black}sum rate} than \ac{MIMO}-\ac{OMA}, even when there are multiple users in a cluster. Furthermore, for \ac{MIMO}-\ac{NOMA}, the two user case always has a larger {\color{black}sum rate} when compared with the three users case, which matches the finding that when more users are admitted into a cluster, a lower {\color{black}sum rate} is obtained.   

\begin{figure}
\centering
\includegraphics[width=0.8\textwidth]{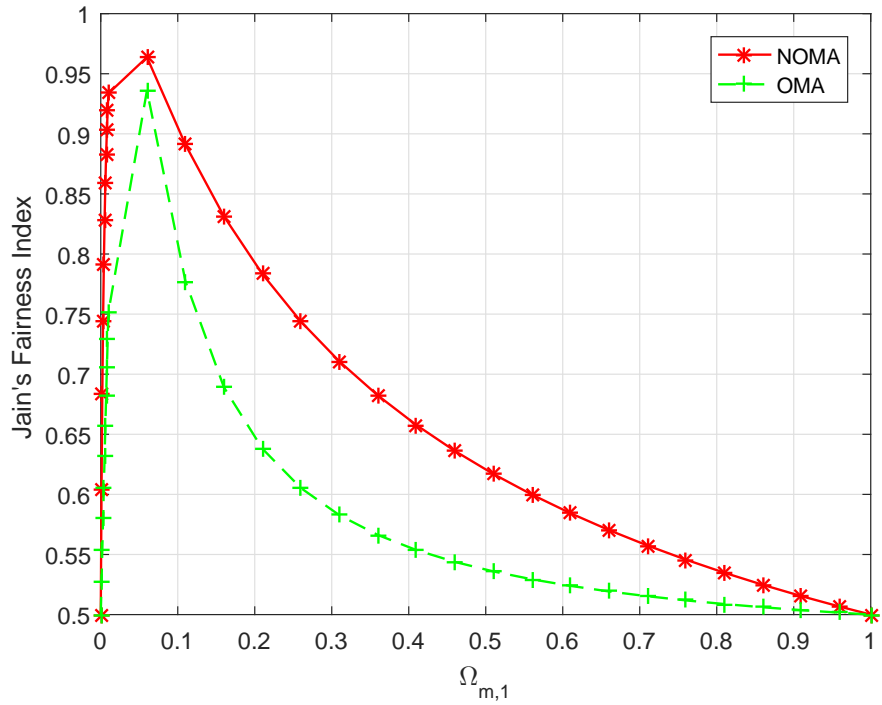}
\caption{Fairness comparison between MIMO-NOMA and MIMO-OMA for two users as the power coefficient varies.}
\end{figure}

\begin{figure}
\centering
\includegraphics[width=0.8\textwidth]{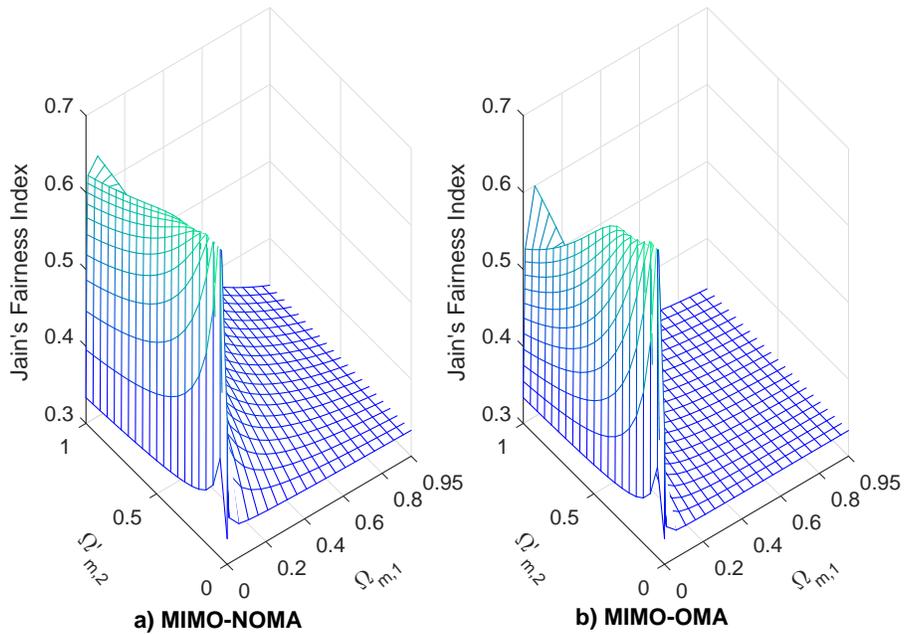}
\caption{Fairness comparison between MIMO-NOMA and MIMO-OMA for three users as the power coefficients vary.}
\end{figure}

Further, Fig. 2 generalizes the case for three users from \text{Fig. 1}, since now an arbitrary power split is provided for all three users. Thus, a three-dimensional figure is displayed, {\color{black}in which the y-axis scaled by $1-\Omega_{m,1}$ represents the power coefficient of the second user, i.e., $\Omega_{m,2}=\Omega_{m,2}^{'} (1-\Omega_{m,1})$.\footnote[3]{Note that in Fig. 2, $\Omega_{m,1}$ does not reach 1. The case of $\Omega_{m,1}=1$ can be seen in Fig. 1, when the sum rates for NOMA and OMA are the same.}} Additionally, the remaining power is allocated to the third user. For both \ac{MIMO}-\ac{NOMA} and \ac{MIMO}-\ac{OMA}, the {\color{black}sum rate} increases significantly with $\Omega_{m,1}$. Meanwhile, when $\Omega_{m,1}$ is fixed, both {\color{black}sum rates} grow gradually with $\Omega_{m,2}$. These again illustrate that when more power is allocated to the user with better channel, a higher {\color{black}sum rate} is achieved. On the other hand, when comparing Figs. 2a) and 2b), it can be seen that \ac{MIMO}-\ac{NOMA} always obtains a higher {\color{black}sum rate} than \ac{MIMO}-\ac{OMA} for any power split among the users, which is in accordance with Theorem 2. Indeed, the maximum gap between \ac{MIMO}-\ac{NOMA} and \ac{MIMO}-\ac{OMA} is 2.04 bps/Hz, which is obtained at the point with $\Omega_{m,1}=0.05,\Omega_{m,2}=0.95$. In this case, only two users are admitted, and this can be explained by the fact that the two user case has a larger sum rate, which is likely to lead to a larger gap. For the two user case, the power allocation coefficients are consistent with the conclusion of Lemma 4, since during the simulation, $\rho|{\mbf{v}_{m,1}^H}\mbf{H}_{m,1}\mbf{p}_m|^2=321$, and thus we have $\Omega_{m,1}=0.053$, which is close to $0.05$. 

\begin{figure}
\centering
\includegraphics[width=0.8\textwidth]{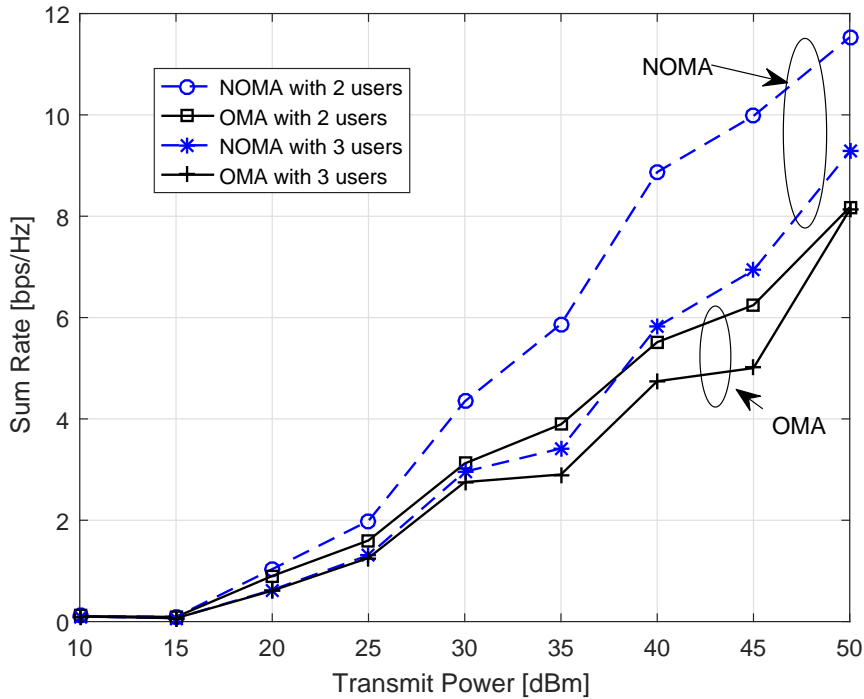}
\caption{{\color{black}Sum rate} for \ac{MIMO-NOMA} and \acs{MIMO-OMA} vs. the transmit power.}
\end{figure} 

\begin{figure}
\centering
\includegraphics[width=0.8\textwidth]{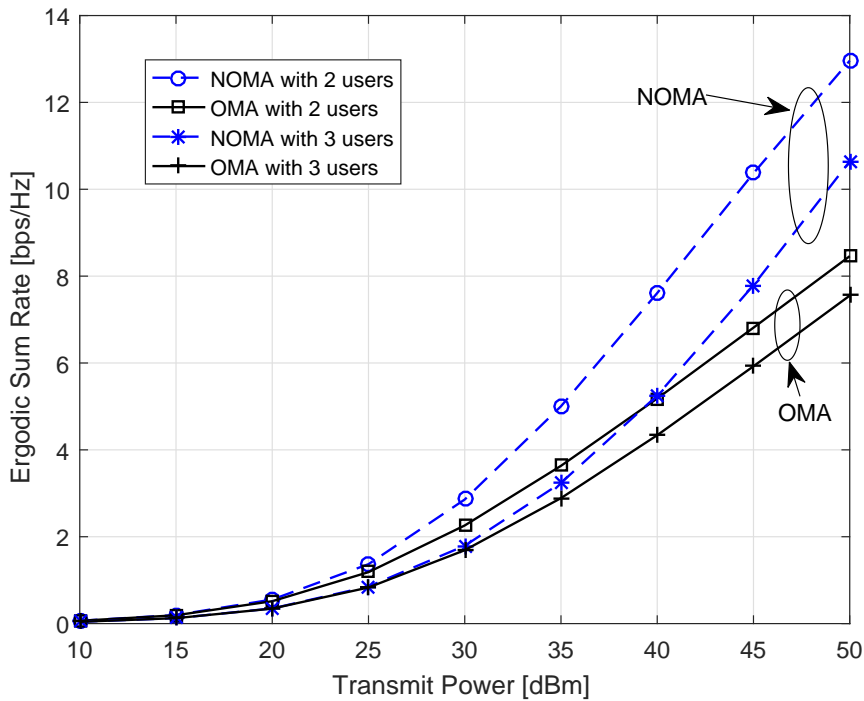}
\caption{{\color{black}Ergodic sum rate} for \ac{MIMO-NOMA} and \acs{MIMO-OMA} vs. the transmit power.}
\end{figure}

{\color{black}Figs. 3 and 4 compare the Jain's fairness index (JFI) \cite{17} of MIMO-NOMA and MIMO-OMA when there are two and three users in a cluster, respectively. Note that $\Omega_{m,2}^{'}$ has the same meaning as in Fig. 2. For both MIMO-NOMA and MIMO-OMA, for the two users case, the JFI first increases with the power coefficient to the first user ($\Omega_{m,1}$). After a certain point, i.e., around 0.1, the JFI decreases as $\Omega_{m,1}$ grows. This trend is expected, as when $\Omega_{m,1}$ is small, increasing its value leads to a more balanced rate distribution between the two users. After the point where the data rate of the first user reaches that of the second user, increasing $\Omega_{m,1}$ results in less fair rate distribution. For the three user case, as shown in Fig. 4, the JFI exhibits the same trend as $\Omega_{m,1}$ varies. When $\Omega_{m,1}$ is fixed, the relationship between JFI and $\Omega_{m,2}$ is more complex, and depends on the specific value of $\Omega_{m,1}$. In all, it can be seen that MIMO-NOMA dominates MIMO-OMA in both cases, which validates that MIMO-NOMA exhibits better fairness when compared with MIMO-OMA.
}

% the idea that \ac{MIMO}-\ac{NOMA} achieves a better performance than \ac{MIMO}-\ac{OMA} when a larger proportion of power is allocated to the user with worse channel condition.    

Figs. 5 and 6 respectively investigate the {\color{black}sum rate} and {\color{black}ergodic sum rate} variation with the transmit power for \ac{MIMO}-\ac{NOMA} and \ac{MIMO}-\ac{OMA}. Although there exists some fluctuation in Fig. 5, due to the variation of the wireless channel, it is still quite clear that the {\color{black}sum rate} of both \ac{MIMO}-\ac{NOMA} and \ac{MIMO}-\ac{OMA} grows with the transmit power. This trend becomes more obvious in Fig. 6, since the ergodic operation reduces the fluctuation of the channel. Moreover, in both two and three user cases, the {\color{black}sum rate} and {\color{black}ergodic sum rate} of \ac{MIMO}-\ac{NOMA} is larger than that of \ac{MIMO}-\ac{OMA}, which further validates our finding in Theorem 2. Meanwhile, as for \ac{MIMO}-\ac{NOMA}, the two user case always has a larger {\color{black}sum rate} and {\color{black}ergodic sum rate} than the three users case, which also verifies our point that as the number of admitted users increases in a cluster, the {\color{black}sum rate} decreases.  

\begin{figure}
\centering
\includegraphics[width=0.8\textwidth]{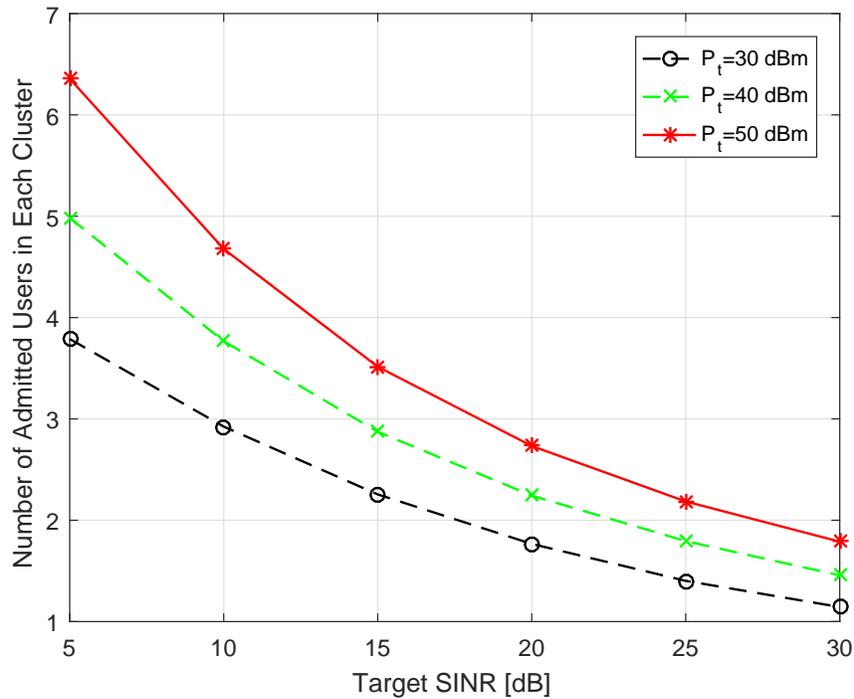}
\caption{Number of admitted users vs. target \ac{SINR}.}
\end{figure}
\begin{figure}
\centering
\includegraphics[width=0.8\textwidth]{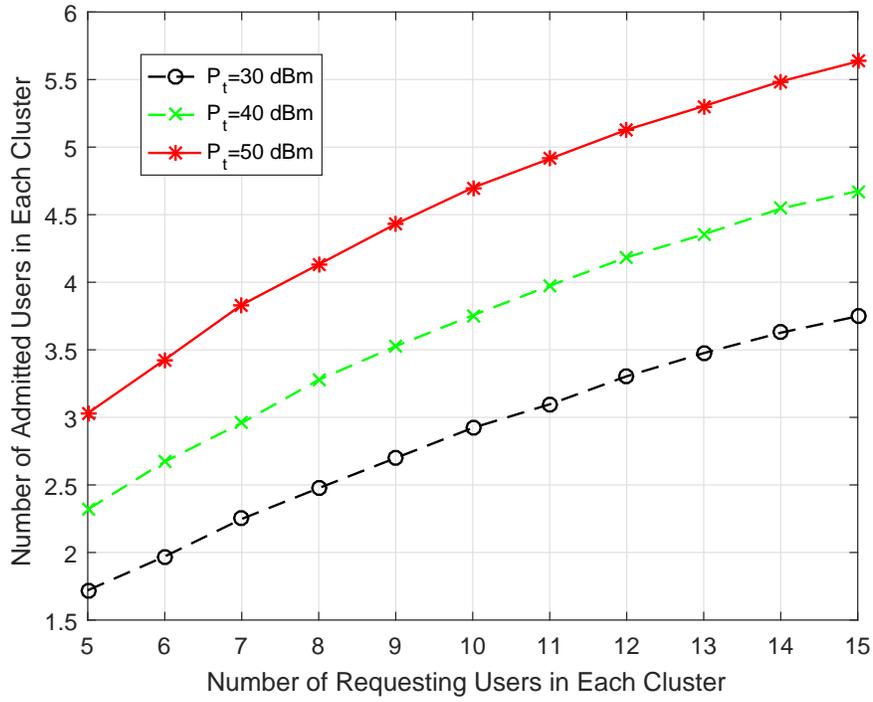}
\caption{Number of admitted users vs. number of requesting users with different transmit power.}
\end{figure} 

\begin{figure}
\centering
\includegraphics[width=0.8\textwidth]{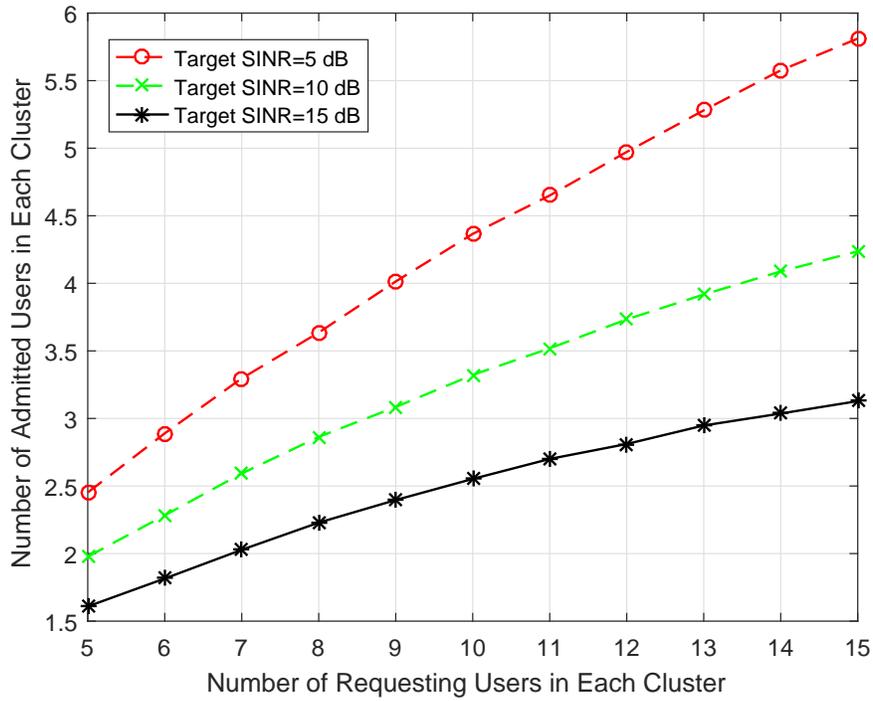}
\caption{Number of admitted users vs. number of requesting users with different target \ac{SINR}.}
\end{figure}

In Figs. 7, 8 and 9, we focus on the performance of the proposed user admission scheme. As shown in Fig. 7, the number of admitted users per cluster declines with the target \ac{SINR} regardless of the transmit power level. This can be easily explained by the fact that as the target \ac{SINR} increases, more power is needed to satisfy each admitted user. Since the total transmit power is fixed, the number of admitted users decreases accordingly. On the other hand, if the total transmit power increases, more users can be admitted, which is verified by the difference in the number of admitted users when the total transmit power is 30 dBm, 40 dBm and \text{50 dBm}, respectively. When the target \ac{SINR} is 5 dB, about 4 users can be admitted into each cluster even when the total transmit power is 30 dBm, which indicates the effectiveness of the proposed user admission scheme. Further, when the total transmit power is 50 dBm, about 6.5 users on average are admitted to each cluster.

\begin{figure}
\centering
\includegraphics[width=0.8\textwidth]{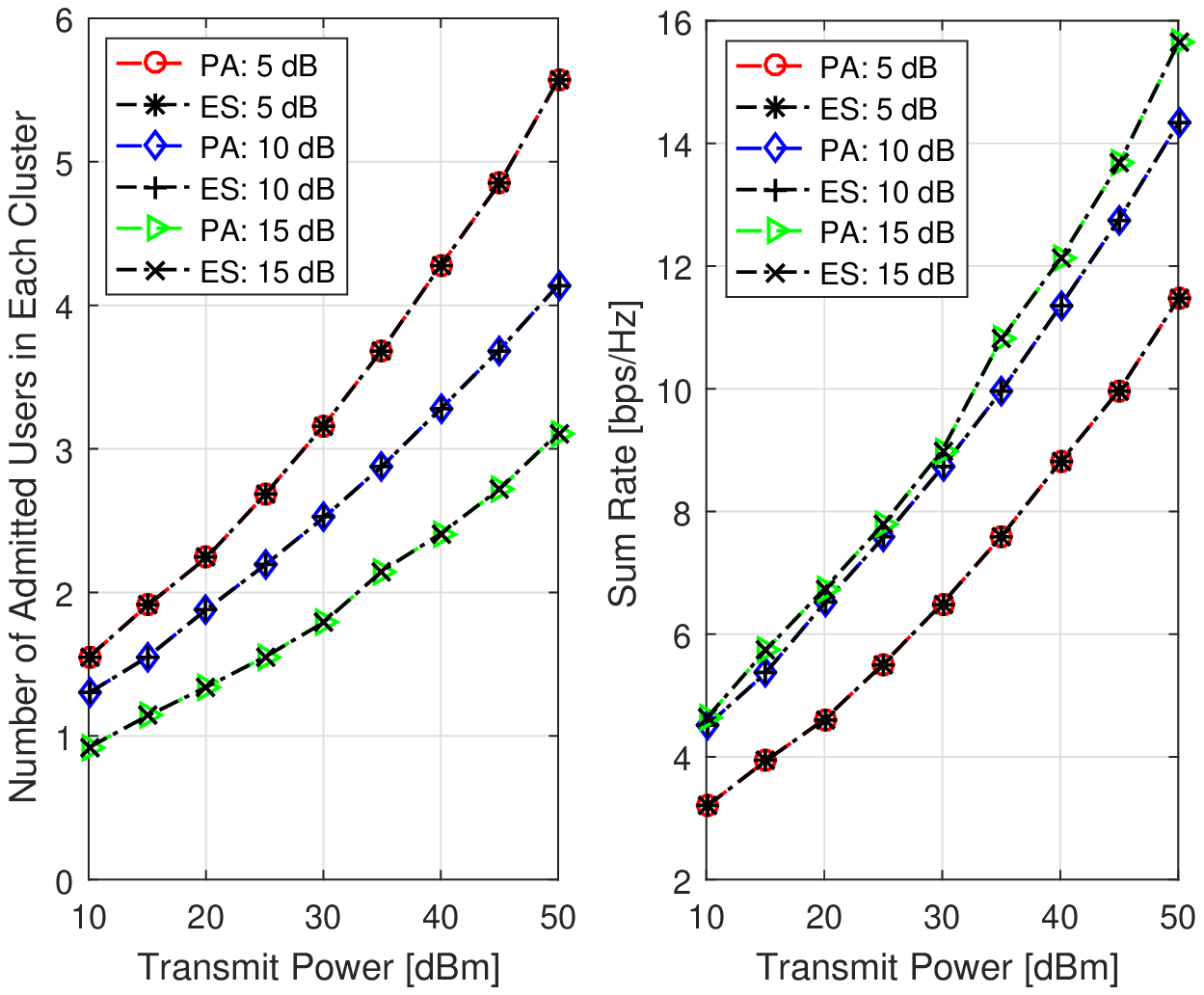}
\caption{Proposed algorithm vs. exhaustive search when the target SINRs of the users are equal.}
\end{figure}

\begin{figure}
\centering
\includegraphics[width=0.8\textwidth]{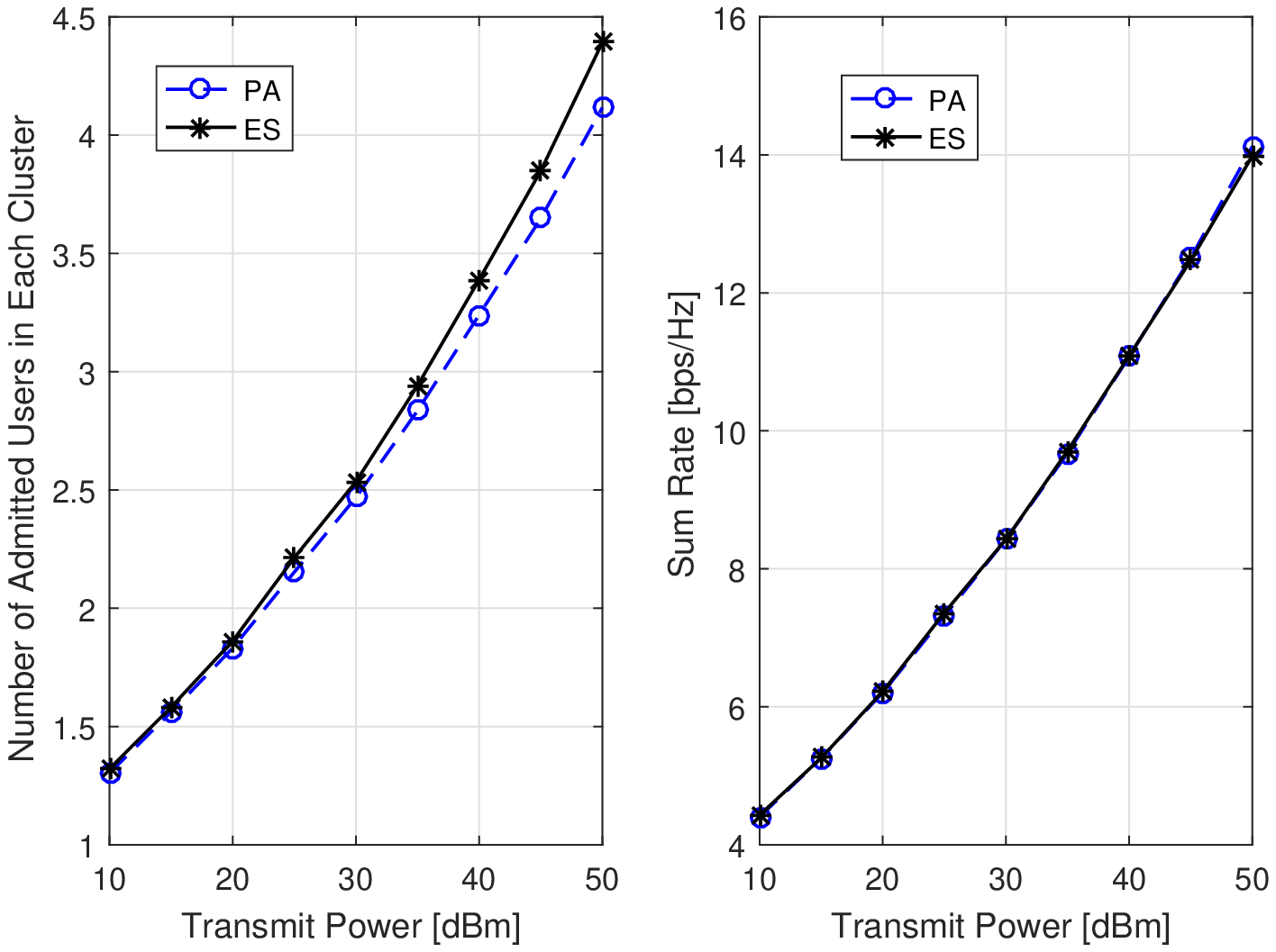}
\caption{Proposed algorithm vs. exhaustive search when the user target SINRs are different.}
\end{figure}

Figs. 8 and 9 illustrate how the number of admitted users per cluster varies with that of the requesting users per cluster. Specifically, Fig. 8 shows results for different transmit powers, while Fig. 9 displays results for different target \ac{SINR}s. Note that the target \ac{SINR} is set to 10 dB in Fig. 8, whereas the total transmit power is set to 35 dBm in Fig. 9. From these figures, it can be observed that the number of admitted users per cluster grows with that of the requesting users. This is due to the fact that with more users requesting admission, more users are likely to have a better channel. According to the proposed user admission scheme, i.e., (\ref{eq:PA2}), less power is required to admit one user when it has a good channel gain. Therefore, more users can be admitted with the same total transmit power. Further, as expected, results in Figs. 8 and 9 show that the number of admitted users per cluster grows with the total transmit power, while it decreases with the target \ac{SINR}, respectively. 

In Figs. 10 and 11, the performance of the proposed algorithm and exhaustive search is compared. Specifically, the exhaustive search is conducted as follows: first, we consider all possible combinations of the users; then, for each combination, we use (\ref{eq:PA2}) to allocate the power coefficient to each user, and decide whether this combination is feasible or not; among all feasible combinations, we select the ones with the largest number of users; lastly, from the selected ones, the one with the highest {\color{black}sum rate} is chosen. In simulations, the number of requesting users is 8, and results are obtained from 1000 trials. Note that PA and ES in the legend represent the proposed algorithm and exhaustive search, respectively. 

In Fig. 10, the target SINR of all users is equal, and the number in the legend represents its value. According to Fig. 10, it can be seen that the performance of the proposed algorithm is the same as the one of the exhaustive search in terms of both {\color{black}sum rate} and number of admitted users for all three target SINRs. In addition, the number of admitted users decreases with the target SINRs, while the {\color{black}sum rate} exhibits an opposite trend. The former can be easily explained, whereas the latter is due to the fact that the increase in the data rate of the admitted users dominates the decrease in the number of admitted users. 

Furthermore, in Fig. 11, the comparison is conducted when the target SINRs are different. Specifically, each user is randomly assigned a target SINR value of 5, 10, or 15 dB. As can be seen from Fig. 11, the exhaustive search achieves better result in terms of the number of admitted users per cluster. However, the gap between the proposed algorithm and exhaustive search is minor. In particular, when the transmit power is 50 dBm, the gap reaches a peak, which is only 0.27. On the other hand, as for the {\color{black}sum rate}, a similar performance is achieved. Additionally, the complexity of exhaustive search is $N!$, while the proposed algorithm has a low complexity, i.e., linear to the number of users per cluster. To conclude, these results verify the effectiveness of the proposed algorithm also when the users' target SINRs are different.

\section{Conclusion}
We have compared the {\color{black}capacity} of \ac{MIMO-NOMA} with that of \acs{MIMO-OMA}, when multiple users are grouped into a cluster. First, we have demonstrated the superiority of \ac{MIMO-NOMA} over \acs{MIMO-OMA} in terms of both {\color{black}sum channel capacity} and {\color{black}ergodic sum capacity}. Furthermore, we have derived the power coefficient value that maximizes the {\color{black}sum rate} gap between \ac{MIMO-NOMA} and \acs{MIMO-OMA}, when there are two users per cluster. {\color{black}Meanwhile, for two and three users per cluster, numerical results also verify that \ac{MIMO-NOMA} dominates \acs{MIMO-OMA} in terms of user fairness.} We have also proved that the more users are admitted to the same cluster, the lower is the achieved {\color{black}sum rate}, which implies a tradeoff between {\color{black}sum rate} and number of admitted users. On this basis, we have proposed a user admission scheme, which achieves optimal results in terms of both {\color{black}sum rate} and number of admitted users when the \ac{SINR} thresholds of the users are equal. When the \ac{SINR} thresholds of the users are different, the proposed scheme still achieves good performance in balancing both criteria. Furthermore, the proposed scheme is of low complexity, i.e., linear in the number of users in each cluster. Finally, the developed analytical results have been validated by simulation results.

\appendices
\section{Proof of Lemma 1}
At the receiver side of user $(m,l)$, the following constraint has to be satisfied in order to implement \ac{SIC} effectively:
\begin{equation}
R_{m,l}^{k} \geq R_{m,k}^{\text{NOMA}}, \forall{k}\in \{l+1,\dots,L \},
\end{equation}
where $R_{m,l}^{k}$ denotes the data rate of user $(m,k)$ achieved at the receiver $(m,l)$, whereas $R_{m,k}^{\text{NOMA}}$ represents the achievable data rate of user $(m,k)$ at its receiver side. Indeed, the above equation guarantees that user $(m,l)$ can remove the interference of those users with worse channel gains, i.e., $(m,l+1), \dots, (m,L)$. According to the order of the effective channel gains, i.e., $|{\mbf{v}_{m,l}^H}\mbf{H}_{m,l}\mbf{p}_m|^2 \geq |{\mbf{v}_{m,k}^H}\mbf{H}_{m,k}\mbf{p}_m|^2, \forall k \geq l$, we have
\begin{equation} \label{eq:PA3}
\begin{split}
R_{m,l}^{k}&= \log_2
\begin{pmatrix}
1+ \frac{\rho \Omega_{m,k} |{\mbf{v}_{m,l}^H}\mbf{H}_{m,l}\mbf{p}_m|^2 }
{1+  \rho \sum_{i=1}^{k-1} \Omega_{m,i} |{\mbf{v}_{m,l}^H}\mbf{H}_{m,l}\mbf{p}_m|^2}
\end{pmatrix} \\ 
&\geq \log_2\begin{pmatrix}
1+ \frac{\rho \Omega_{m,k} |{\mbf{v}_{m,k}^H}\mbf{H}_{m,k}\mbf{p}_m|^2 }
{1+  \rho \sum_{i=1}^{k-1} \Omega_{m,i} |{\mbf{v}_{m,k}^H}\mbf{H}_{m,k}\mbf{p}_m|^2}
\end{pmatrix} \\ 
&=R_{m,k}^{\text{NOMA}}.
\end{split}
\end{equation}

Thus, $R_{m,l}^{k} \geq R_{m,k}^{\text{NOMA}}, \forall{k}\in \{l+1,\dots,L \}$ is always true. Consequently, the use of \ac{SIC} is always guaranteed at the receiver $(m,l)$ owing to the ordering of the effective channel gains, and this puts no extra constraints on the system.

\section{Proof of Theorem 1}
For simplicity of notation, let $K_l=\rho \Omega_{m,l} |{\mbf{v}_{m,l}^H}\mbf{H}_{m,l}\mbf{p}_m|^2, \: l\in \{1, \dots, L \}$. 
Theorem 1 can be proved via mathematical induction, and the hypothesis is
\begin{equation}
{\color{black}S_{m,L_1}^{\text{OMA}} } \leq (\sum_{l=1}^{L_1} \lambda_{m,l})
 \log_2 (1+ \frac{\sum_{l=1}^{L_1}K_l}{\sum_{l=1}^{L_1} \lambda_{m,l} }   ),
\end{equation} 
where ${\color{black}S_{m,L_1}^{\text{OMA}} }$ represents the sum rate for the first $L_1$ users, $L_1 \in \{1, \dots, L \}$. Obviously, the first user satisfies the hypothesis, since ${\color{black}S_{m,1}^{\text{OMA}} }=R_{m,1}^{\text{OMA}}=\lambda_{m,1}
 \log_2 (1+ \frac{K_1}{ \lambda_{m,1} } )$. 
 
Then, let us consider the case of $L_2=L_1+1$, and we have
\begin{equation} \label{eq1}
\begin{split}
&{\color{black}S_{m,L_2}^{\text{OMA}} } \\
&={\color{black}S_{m,L_1}^{\text{OMA}} }+  \lambda_{m,L_2} \log_2 (1+ \frac{K_{L_2}}{ \lambda_{m,L_2}} )  \\ 
&\leq (\sum_{l=1}^{L_1} \lambda_{m,l})  \log_2 (1+ \frac{\sum_{l=1}^{L_1}K_l}{\sum_{l=1}^{L_1} \lambda_{m,l} }   )
+\lambda_{m,L_2} \log_2 (1+ \frac{K_{L_2}}{ \lambda_{m,L_2}} ) \\ 
&=( \sum_{l=1}^{L_2} \lambda_{m,l} ) 
\bigg[ \frac{ \sum_{l=1}^{L_1} \lambda_{m,l} } { \sum_{l=1}^{L_2} \lambda_{m,l}} 
 \log_2 (1+ \frac{\sum_{l=1}^{L_1}K_l }{\sum_{l=1}^{L_2} \lambda_{m,l} } \frac{\sum_{l=1}^{L_2} \lambda_{m,l}}{\sum_{l=1}^{L_1} \lambda_{m,l}}  ) \\
\:&+\frac{\lambda_{m,L_2}}{(\sum_{l=1}^{L_2} \lambda_{m,l})}  \log_2 (1+ {\color{black}\frac{K_{L_2}}{ \sum_{l=1}^{L_2} \lambda_{m,l} } }  \frac{\sum_{l=1}^{L_2} \lambda_{m,l}}{\lambda_{m,L_2}} ) \bigg].
\end{split}
\end{equation}

Let $\lambda=\frac{ \sum_{l=1}^{L_1} \lambda_{m,l} } { \sum_{l=1}^{L_2} \lambda_{m,l}}$, then $1-\lambda=\frac{\lambda_{m,L_2}}{(\sum_{l=1}^{L_2} \lambda_{m,l})}$. In addition, let $K_1'=\frac{\sum_{l=1}^{L_1}K_l }{\sum_{l=1}^{L_2} \lambda_{m,l} }$ and $K_2'={\color{black}\frac{K_{L_2}}{ \sum_{l=1}^{L_2} \lambda_{m,l} } }$. The polynomial in the bracket can be reformulated as $\lambda \log_2 (1+ \frac{K_1'} {\lambda} )+
(1-\lambda) \log_2 ( 1+ \frac{K_2' } {1-\lambda})$, which has the same form as \text{[15, eq. (12)]}. According to Lemma 2, it can be written as $\log_2(1+\frac{\sum_{l=1}^{L_2}K_l }{\sum_{l=1}^{L_2} \lambda_{m,l} })$, satisfying $ \frac{\sum_{l=1}^{L_1}K_l }{{\color{black} \sum_{l=1}^{L_1} \lambda_{m,l} } }=\frac{K_{L_2}}{ \lambda_{m,L_2}}$. Thus, we have ${\color{black}S_{m,L_2}^{\text{OMA}} }\leq( \sum_{l=1}^{L_2} \lambda_{m,l} )\log_2(1+\frac{\sum_{l=1}^{L_2}K_l }{\sum_{l=1}^{L_2} \lambda_{m,l} })$, which also fits the hypothesis.

Lastly, we consider the case for $L$ users. Due to $\sum_{l=1}^{L}\lambda_{m,l}=1$, we have ${\color{black}S_{m,L}^{\text{OMA}} } \leq \log_2(1+\sum_{l=1}^{L}K_l)=\log_2(1+\sum_{l=1}^{L}\rho \Omega_{m,l} |{\mbf{v}_{m,l}^H}\mbf{H}_{m,l}\mbf{p}_m|^2)$. Here Theorem 1 is proved. Moreover, it is easy to conclude that the equality is achieved when $\frac{ \Omega_{m,1} |{\mbf{v}_{m,1}^H}\mbf{H}_{m,1}\mbf{p}_m|^2 } 
{\lambda_{m,1}}= \dots =  \frac{\Omega_{m,l} |{\mbf{v}_{m,L}^H}\mbf{H}_{m,L}\mbf{p}_m|^2 } 
{\lambda_{m,L}} $. Correspondingly, we have ${\lambda_{m,l}=\frac{ \Omega_{m,l} |{\mbf{v}_{m,l}^H}\mbf{H}_{m,l}\mbf{p}_m|^2 } { \sum_{1}^{L}  \Omega_{m,l} |{\mbf{v}_{m,l}^H}\mbf{H}_{m,l}\mbf{p}_m|^2} }, \forall \: l\in\{1,\dots,L\}$.

\section{Proof of Lemma 3}
According to inequality (\ref{eq:order}), we have $\rho \Omega_{m,k} |{\mbf{v}_{m,k}^H}\mbf{H}_{m,k}\mbf{p}_m|^2 \geq \rho \Omega_{m,k} |{\mbf{v}_{m,l}^H}\mbf{H}_{m,l}\mbf{p}_m|^2, \forall {k \leq l}$. Consequently, it can be concluded that
\begin{equation} \label{eq:channel}
\frac{1+ \rho \sum_{k=1}^{l}  \Omega_{m,k} |{\mbf{v}_{m,k}^H}\mbf{H}_{m,k}\mbf{p}_m|^2 }{1+  \rho \sum_{k=1}^{l} \Omega_{m,k} |{\mbf{v}_{m,l}^H}\mbf{H}_{m,l}\mbf{p}_m|^2} \geq 1, l \in \{1, \dots, L \}. 
\end{equation}

Further, the above equation can be used to obtain the lower bound for the {\color{black}sum rate} for \ac{MIMO}-\ac{NOMA} via mathematical induction, and the hypothesis is that the sum rate for the first $l$ users, denoted as ${\color{black}S_{m,l}^{\text{NOMA}} }$ is bounded by 
\begin{equation} \label{eq:35}
{\color{black}S_{m,l}^{\text{NOMA}} } \geq \log_2(1+ \rho \sum_{k=1}^{l}  \Omega_{m,k} |{\mbf{v}_{m,k}^H}\mbf{H}_{m,k}\mbf{p}_m|^2  ).
\end{equation}

Clearly, the first user satisfies (\ref{eq:35}), since ${\color{black}S_{m,1}^{\text{NOMA}} }=R_{m,1}^{\text{NOMA}}=\log_2(1+ \rho \Omega_{m,1} |{\mbf{v}_{m,1}^H}\mbf{H}_{m,1}\mbf{p}_m|^2) \geq \log_2(1+ \rho \Omega_{m,1} |{\mbf{v}_{m,1}^H}\mbf{H}_{m,1}\mbf{p}_m|^2)$. 

Next, the case for $l+1$ users is proved as follows:
%\begin{equation} \label{eq:Lemma4}
%\begin{split}
%&\sum_{k=1}^{l+1} R_{m,k}^{\text{NOMA}} \\
%& = \sum_{k=1}^{l} R_{m,k}^{\text{NOMA}}+ R_{m,l+1}^{\text{NOMA}} \\ 
%& \geq \log_2( 1+ \rho \sum_{k=1}^{l}  \Omega_{m,k} |{\mbf{v}_{m,k}^H}\mbf{H}_{m,k}\mbf{p}_m|^2) \\ 
%&+ \log_2 (1+ \frac{\rho \Omega_{m,l+1} |{\mbf{v}_{m,l+1}^H}\mbf{H}_{m,l+1}\mbf{p}_m|^2 }
%{1+  \rho \sum_{k=1}^{l} \Omega_{m,k} |{\mbf{v}_{m,l+1}^H}\mbf{H}_{m,l+1}\mbf{p}_m|^2 }) \\ 
%& = \log_2( 1+ \rho \sum_{k=1}^{l}  \Omega_{m,k} |{\mbf{v}_{m,k}^H}\mbf{H}_{m,k}\mbf{p}_m|^2  \\ 
%&+  \frac{\rho \Omega_{m,l+1} |{\mbf{v}_{m,l+1}^H}\mbf{H}_{m,l+1}\mbf{p}_m|^2  (1+ \rho \sum_{k=1}^{l}  \Omega_{m,k} |{\mbf{v}_{m,k}^H}\mbf{H}_{m,k}\mbf{p}_m|^2)}
%{1+  \rho \sum_{k=1}^{l} \Omega_{m,k} |{\mbf{v}_{m,l+1}^H}\mbf{H}_{m,l+1}\mbf{p}_m|^2 }) \\ 
%& \geq \log_2( 1+ \rho \sum_{k=1}^{l}  \Omega_{m,k} |{\mbf{v}_{m,k}^H}\mbf{H}_{m,k}\mbf{p}_m|^2 \\
%&+ \rho \Omega_{m,l+1} |{\mbf{v}_{m,l+1}^H}\mbf{H}_{m,l+1}\mbf{p}_m|^2 ),
%\end{split}
%\end{equation}
\begin{equation} \label{eq:Lemma4}
\begin{split}
{\color{black}S_{m,l+1}^{\text{NOMA}} }
&= {\color{black}S_{m,l}^{\text{NOMA}} }+ R_{m,l+1}^{\text{NOMA}} \\ 
& \geq \log_2( 1+ \rho \sum_{k=1}^{l}  \Omega_{m,k} |{\mbf{v}_{m,k}^H}\mbf{H}_{m,k}\mbf{p}_m|^2) \\ 
&+ \log_2 (1+ \frac{\rho \Omega_{m,l+1} |{\mbf{v}_{m,l+1}^H}\mbf{H}_{m,l+1}\mbf{p}_m|^2 }
{1+  \rho \sum_{k=1}^{l} \Omega_{m,k} |{\mbf{v}_{m,l+1}^H}\mbf{H}_{m,l+1}\mbf{p}_m|^2 }) \\ 
& = \log_2( 1+ \rho \sum_{k=1}^{l}  \Omega_{m,k} |{\mbf{v}_{m,k}^H}\mbf{H}_{m,k}\mbf{p}_m|^2 \\
& + \rho \Omega_{m,l+1} |{\mbf{v}_{m,l+1}^H}\mbf{H}_{m,l+1}\mbf{p}_m|^2 \\
&\times \frac{  (1+ \rho \sum_{k=1}^{l}  \Omega_{m,k} |{\mbf{v}_{m,k}^H}\mbf{H}_{m,k}\mbf{p}_m|^2)}
{1+  \rho \sum_{k=1}^{l} \Omega_{m,k} |{\mbf{v}_{m,l+1}^H}\mbf{H}_{m,l+1}\mbf{p}_m|^2 }) \\ 
& \geq \log_2( 1+ \rho \sum_{k=1}^{l+1}  \Omega_{m,k} |{\mbf{v}_{m,k}^H}\mbf{H}_{m,k}\mbf{p}_m|^2),
\end{split}
\end{equation}
where the last inequality comes from (\ref{eq:channel}). 

Thus, when all $L$ users are considered, we have ${\color{black}S_{m,L}^{\text{NOMA}} } \geq \log_2(1+ \rho \sum_{k=1}^{L}  \Omega_{m,k} |{\mbf{v}_{m,k}^H}\mbf{H}_{m,k}\mbf{p}_m|^2  )$. Hence, Lemma 3 is proved. 

\section{Proof of Theorem 3}
Consider the case in which only $l$ users can be admitted to the $m$th cluster when employing the proposed user admission scheme. Suppose there exists an alternate scheme, which also admits $l$ users, denoted as $a_1,a_2,\dots,a_{l}$. Theorem 3 can be proved through contradiction.

Specifically, the proof consists of two steps: 1) it is shown that the sum power required by the alternate scheme always exceeds that of the proposed scheme; and 2) based on (1), assume that the alternate scheme can admit an extra user, this user should also be admitted by the proposed scheme, which conflicts with the proposition that only $l$ users can be admitted by the proposed scheme. Consequently, no other scheme can admit a larger number of users than the proposed one.

Step 1: The power coefficients of the proposed scheme and the alternate one are denoted as  $\Omega_1,\Omega_2,\dots,\Omega_{l}$, and $\Omega_{a_1},\Omega_{a_2},\dots,\Omega_{a_{l}}$, respectively. For notational simplicity, let $G_{k}=|{\mbf{v}_{k}^H}\mbf{H}_{k}\mbf{p}|^2, k \in \{ 1 \:,2 \: ,\dots,l \}$, and $G_{a_{k}}=|{\mbf{v}_{a_{k}}^H}\mbf{H}_{a_{k}}\mbf{p}|^2, a_{k} \in \{ a_1,a_2,\dots,a_{l} \}$. Without loss of generality, the admitted $l$ users for the alternate scheme are also ranked in a descending order according to their channel gains, i.e., $G_{a_1} \geq  \dots  \geq G_{a_{l}}$. Thus, it can be easily observed that $G_{a_{k}} \leq G_{k}$, since $k \leq a_{k}$ due to the channel order and user admission order of both schemes. Moreover, according to (\ref{eq:PA}) and (\ref{eq:PA2}), we have $\Omega_{a_k} \geq \Gamma_{a_k} \sum_{i=1}^{k-1}{\Omega_{a_i}} +\frac{\Gamma_{a_k}}{\rho G_{a_{k}} }$, and  $\Omega_k = \Gamma_k \sum_{i=1}^{k-1}{\Omega_i} +\frac{\Gamma_k}{\rho G_k }$, respectively. After some algebraic manipulations, the sums of the power coefficients for the proposed scheme and the alternate one can be expressed as 
\begin{subequations} \label{eq:user1}
\begin{align}
\Psi&=\sum_{k=1 }^{l}{\frac{\Gamma{k} }{\rho G_k} \prod_{i=k+1}^{l} (\Gamma_{i}+1) } \\
\Psi_{a}& \geq \sum_{a_k=1 }^{a_l}{\frac{\Gamma{a_k} }{\rho G_{a_k}} \prod_{a_i=a_k+1}^{a_l} (\Gamma_{a_i}+1) },
\end{align}
\end{subequations}
where $\Psi$ and $\Psi_{a}$ denote the sums of the power coefficients for the proposed scheme and the alternate one, respectively.

By using (\ref{eq:SINRvary}a), (\ref{eq:SINRvary}b) and $G_{a_{k}} \leq G_{k}$, it can be easily obtained that $\frac{\Gamma{k} }{G_k} \leq \frac{\Gamma{a_k} }{G_{a_k}}$, and $\prod_{i=k+1}^{l} (\Gamma_{i}+1)  \leq \prod_{a_i=a_k+1}^{a_l} (\Gamma_{a_i}+1) $. Thus, $\Psi \leq \Psi_{a}$, which means that to admit the same number of users, the proposed scheme requires the minimum power. 

Step 2: Suppose the alternate scheme can admit an extra user, $a_{l+1}$, whose power coefficient and channel gain are denoted as $\Omega_{a_{l+1}}$ and $G_{a_{l+1}}$, respectively. According to (\ref{eq:PA}) and (\ref{eq:user}b), we have $\Omega_{a_{l+1}} \geq \Gamma_{a_{l+1}} \Psi_{a} +\frac{\Gamma_{a_{l+1}}}{\rho G_{a_{{l+1}}} } $, which must satisfy $\Omega_{a_{l+1}}+\Psi_{a}\leq 1$. On this basis, it is easy to verify that user $a_{l+1}$ can also be admitted by the proposed scheme, since $\Omega_{a_{l+1}}'+\Psi\leq \Omega_{a_{l+1}}+\Psi_{a}\leq 1$, where $\Omega_{a_{l+1}}'=\Gamma_{a_{l+1}} \Psi +\frac{\Gamma_{a_{l+1}}}{\rho G_{a_{{l+1}}} }$ denotes the power coefficient of user $a_{l+1}$ under the proposed scheme. Clearly, this conflicts with the proposition that only $l$ users can be admitted by the proposed scheme.  

%}}

\bibliographystyle{IEEEtran}
\bibliography{IEEEabrv,conf_short,jour_short,mybibfile}

\end{document}